\documentclass[natbib,smallextended]{svjour3}

\usepackage{graphicx}
\usepackage{amssymb} \usepackage{epstopdf}
\usepackage{hyperref}
\usepackage{url}
\usepackage{comment}
\usepackage{float}
\usepackage{color}
\usepackage{multirow}
\usepackage[most]{tcolorbox}
\usepackage{tabularx}
\usepackage{balance}
\usepackage{changepage} 
\usepackage{placeins}
\usepackage{longtable}
\usepackage[normalem]{ulem}
\journalname{Empirical Software Engineering}

\begin{document}

\title{Peer Code Review in Research Software Development: The Research Software Engineer Perspective}

\author{Md Ariful Islam Malik \and
        Jeffrey C. Carver         \and
        Nasir U. Eisty }

\institute{M. Malik \at
              Department of Computer Science, University of Alabama,
             Tuscaloosa, AL, USA\\
              \email{mmalik3@crimson.ua.edu}
           \and
           J. C. Carver \at
              Department of Computer Science,
              University of Alabama,
             Tuscaloosa, AL, USA\\
              \email{carver@cs.ua.edu}
            \and
            N. U. Eisty \at
            Department of EECS,
             University of Tennessee,
            Knoxville, TN, USA \\
            \email{neisty@utk.edu}
}

\date{Received: date / Accepted: date}

\newcommand*{\RQa}[0]{\textbf{RQ1: How do research software engineers perform peer code review?}}
\newcommand*{\RQb}[0]{\textbf{RQ2: What effect does peer code review have on research software?}}
\newcommand*{\RQc}[0]{\textbf{RQ3: What difficulties do research software engineers face with peer code review?}}
\newcommand*{\RQd}[0]{\textbf{RQ4: What improvements to the peer code review process do research software engineers need?}}

\maketitle

\begin{abstract}
\textit{Background}: Research software is crucial for enabling research discoveries and supporting data analysis, simulation, and interpretation across domains. 
However, evolving requirements, complex inputs, and legacy dependencies hinder the software quality and maintainability. 
While peer code review can improve software quality, its adoption by research software engineers (RSEs) remains unexplored.
\textit{Aims}: This study explores RSE perspectives on peer code review, focusing on their practices, challenges, and potential improvements. 
Building on prior work, it aims to uncover how RSEs’ insights differ from those of other research software developers and identify factors that can enhance code review adoption in this domain.
\textit{Method}: We surveyed RSEs to gather insights into their perspectives on peer code review. 
The survey design aligned with previous research to enable comparative analysis while including additional questions tailored to RSEs. 
\textit{Results}: We received 61 valid responses from the survey. 
The findings align with prior research while uncovering unique insights about the challenges and practices of RSEs compared to broader developer groups. 
\textit{Conclusions}: 
Peer code review is vital in improving research software's quality, maintainability, and reliability. 
Despite the unique challenges RSEs face, addressing these through structured processes, improved tools, and targeted training can enhance peer review adoption and effectiveness in research software development.

 \keywords{Code review \and Survey \and Research Software \and Software Engineering}
\end{abstract}

\section{Introduction}  
\label{sec:introduction}

\textbf{Research software} is software developed specifically to facilitate or conduct research, analysis, or discovery~\citep{sochat2022research, combemale2023research}. 
It includes tools, applications, libraries, and systems that enable researchers to simulate, analyze, model, or interpret data and complex phenomena. 
As research across disciplines increasingly relies on computational methods, research software has become integral to addressing complex problems, ensuring reproducibility, and accelerating scientific discovery~\citep{combemale2023research, Barker2020}. 
The importance and prevalence of research software make it critical that we better understand how to appropriately check its quality. Low quality research software can lead researchers to draw incorrect conclusions.

Code review is a well-established software quality improvement practice. 
For research software, code review plays a crucial role in overcoming the challenges associated with verification and correctness, yielding multiple benefits. For example, code reviews not only highlight functional defects and suggest code improvements but also convey feedback clearly and respectfully, both aspects significantly influence the comment’s usefulness~\citep{turzo2024makes}. 
Despite the benefits of peer code review, developers of research software do not yet consistently use it~\citep{article159}.

To better understand how to increase the use of peer code review in research software development, we previously conducted a study with interviews followed by a survey of people who develop research software~\citep{Eisty-Carver:2022}.
The primary goal of that study was to gain a deeper understanding of the practices, difficulties, impacts, and potential areas for improvement when using peer code review for research software. 
The results showed that while developers of research software found code review beneficial, they used it informally and less frequently than developers of industrial and open-source projects.
Some of the identified benefits include improved code quality and increased knowledge sharing.
However, the results also showed some negative aspects to peer code review, including that it is time-consuming and that developers misunderstand the criticism of their code.
These results motivate the need for further study of peer code review in research software.

The previous study focused on the broader community of researchers, graduate students, postdocs, and faculty members who develop research software (generally referred to as \textit{research software developers}), many of whom lack formal software engineering training.
While this approach has yielded viable research software, it presents several software challenges, including reproducibility, maintainability, usability, and sustainability. 
To address these issues, the role of \textit{Research Software Engineer (RSE)} has emerged.
RSEs are professionals who combine software development expertise with a deep understanding of the research domain~\citep{ANL_RSE_Career}. 
They use software engineering practices such as testing, documentation, version control, and maintainable design to ensure that research software is reliable, sustainable, and reusable. In addition to programming skills, RSEs have domain-specific knowledge that allows them to work closely with researchers to translate scientific requirements into robust computational tools.

The prior study focused on the broader community of research software developers, rather than specifically on RSEs.
Because of their more detailed knowledge of software engineering practices and processes, we expect that RSEs will have a different experience with peer code review.
If we attempt to directly apply the findings from our previous survey to this more experienced subset of developers without first assessing whether those results hold for this new population, we risk wasting effort or even reducing their effectiveness.
As the prominence of RSEs increases in the research software process, 
it is critical to understand whether their needs related to code review differ from those of research software developers.

Therefore, we replicated the previous study, with a specific focus on RSEs, to determine whether there were any differences in results with this new population and to gain a deeper understanding of the use of peer code review in research software. 
Understanding the perceptions of RSEs is critical because their influence directly affects the effectiveness with which these practices are deployed into research software projects. 
Insights from RSEs can help identify practical barriers to the broader adoption of peer code review, which contributes to more sustainable, reliable research software.

As this study builds on our prior work, the research questions are the same, however, with a new focus on research software engineers:

\begin{itemize}
    \item \RQa 
    \item \RQb
    \item \RQc
    \item \RQd
\end{itemize}

 \section{Background}
\label{sec:background}
In this section, we define research software, compare it with commercial software, describe the role of Research Software Engineers (RSEs), and explain how RSEs differ from software engineers.

\subsection{Research Software}
To clarify the differences between research software and traditional software, we start with a definition of research software, followed by a comparison of research software and traditional, commercial software.

\subsubsection{Definition}
Research software is designed specifically to support or conduct research. 
Specifically, ``research Software includes source code files, algorithms, scripts, computational workflows, and executables that were created during the research process or for a research purpose''~\citep{gruenpeter-etal}.
It frequently includes scientific models, methods, and data analysis tools necessary for scholarly research. 
Compared with general-purpose software, it is closely linked to research activities and plays a direct role in producing, processing, or analyzing data or models to answer scientific questions or generate publishable results~\citep{1,2,3}. 
Research software is recognized as an essential aspect of research, requiring proper management, credit, and sustainability measures to ensure the validity and reproducibility of scientific findings~\citep{4,5}. 
In summary, research software is created or adapted to facilitate research. Its development, maintenance, and recognition are central to modern scientific practice.

\subsubsection{Research Software vs Commercial Software} 
Research software and commercial software have different goals, development approaches, and distribution models. 
Because research software is an integral part of the research process in various research domains, it is often highly specialized, rapidly changing, and may be less polished than commercial software~\citep{7,8,9}. 
In contrast, commercial software prioritizes usability, reliability, and maintainability, with dedicated support and documentation resources~\citep{4, 10}). 
Key differences between research software and commercial software include~\citep{3,4,8,9}:
\begin{itemize}
    \item \textbf{training of developers} - often informal for research software vs. typically more formal for commercial software
    \item \textbf{finances} - research software frequently uses unreliable grant funding vs. commercial software, which often uses established business models 
    \item \textbf{users} - often internal users or other researchers for research software vs. customers for commercial software
    \item \textbf{testing} - research software is often complex and rapidly evolving making testing more difficult than for commercial software that can use more established testing protocols
\end{itemize}
These differences result in differences in sustainability and quality. 
Overall, research software aims to advance scientific discovery, while commercial software prioritizes stability, scalability, and widespread use.

\subsection{Research Software Engineers (RSEs)} 
To clarify the differences between RSEs and traditional software engineers, this section first defines RSEs and then compares the two roles.

\subsubsection{Definition}
An RSE is a professional who combines software development expertise with a knowledge of research practices, working at the intersection of computing and scientific inquiry. 
The role emerged to describe individuals in the research community who focus on developing, maintaining, and supporting software that enables or advances research, rather than conducting research itself or working solely as traditional software engineers. 
RSEs may perform tasks ranging from writing code for scientific models and data analysis to ensuring software reliability, reproducibility, and sustainability within research projects. 
Their responsibilities and required competencies can vary significantly depending on the institutional context, with some roles resembling those of academic researchers and others resembling those of industry software engineers. 
Fundamentally, RSEs bridge the gap between domain scientists and software engineers, ensuring that research software meets high quality and usability standards to support scientific discovery~\citep{1,11}.

\subsubsection{RSEs vs Software Engineers}
RSEs and Software Engineers (SEs) share foundational software development skills, but their roles and priorities differ significantly. 
RSEs operate at the intersection of research and software engineering, focusing on creating, maintaining, and supporting software that directly enables scientific research and discovery. 
They often collaborate closely with researchers, adapting software to evolving scientific questions and ensuring the reproducibility, reliability, and sustainability of research outputs. 
In contrast, SEs in industry typically develop software for commercial or organizational use, prioritizing scalability, user experience, and business requirements, often within well-defined project scopes and established engineering processes. 
While SEs may specialize in areas like web development, systems engineering, or application design, RSEs require a broader understanding of scientific domains and research methodologies, frequently working in multidisciplinary teams. 
RSEs may also take on roles in project management, training, and mentoring within academic or research settings, reflecting their hybrid position between research and technical development. 
The boundaries between these roles can blur, especially in institutions where RSEs’ work closely resembles that of SEs.
However, the core distinction lies in RSEs’ commitment to advancing research through tailored software solutions~\citep{1,11,34}.

\section{Methodology}
\label{sec:Methodology}
Our previous work provided the foundation for this study. 
We used a survey to gain more focused insights from our target population of RSEs, as opposed to the broader range of software developers surveyed in the previous study.
This change helps us understand how the two groups differ in their application of peer code review. 
To facilitate this comparison, we reused most of the survey questions, allowing us to identify and analyze any differences in perspectives between the two populations. 
In some cases, we introduced new questions to better understand the population and their insights on the effectiveness of code review in research software.
This section outlines our survey questions, the distribution method, and the data analysis approach.

\subsection{Survey Design}
We reused the original survey questions to answer the research questions and enable comparison of the results to those from the previous study.
In some areas, we added new questions specific to the new population of RSEs. 
Figures~\ref{fig_survey_questions} and~\ref{fig_survey_questions2} list the survey questions.
We preface the new questions with a $\bigstar$.

\subsection{Data Collection}
The goal of our data collection process was to reach individuals actively working as research software engineers (RSEs) to gather their experiences and perspectives on peer code review. 
We focused on this group to gain a deeper understanding of code review in the context of research software, in contrast to the broader population of research software developers studied in our previous work.
We created the survey using Qualtrics\footnote{https://www.qualtrics.com/}. 
To reach our target audience, we used several solicitation methods aimed at communities associated with research software and RSEs: 
\begin{itemize}
    \item Better Scientific Software (BSSw)\footnote{\url{https://bssw.io/}}: A community that promotes best practices in scientific software development, especially in high-performance computing. BSSw included the survey in their newsletter and distributed it to relevant email lists.  
    \item Chan Zuckerberg Initiative (CZI)\footnote{\url{https://chanzuckerberg.com/}}: An organization that funds open science and infrastructure projects. CZI included the survey in their community newsletter.  
    \item US-RSE\footnote{\url{https://us-rse.org/}} and the Society of Research Software Engineering\footnote{\url{https://society-rse.org/}}: Two professional associations that support the development and recognition of RSE roles. We posted the survey in relevant Slack channels to reach actively working RSEs directly.  
    \item Social Media: We posted the survey on Twitter/X and requested that survey respondents share the survey link among their networks.
\end{itemize}

Because of the nature of our distribution methods, we cannot estimate the total size of the population who received the survey invitation, nor can we calculate a response rate.

To ensure respondent anonymity, we did not collect email addresses or other personally identifiable information (PII).
Since both the current and previous surveys were anonymous, we cannot determine whether any individuals participated in both studies. 
However, in this study, we explicitly focused on participants who identify as research software engineers (RSEs). 
To ensure that only RSEs participated in the current survey, we employed several methods. First, in our survey solicitation, we clearly stated that the study is intended for those who work as research software engineers (RSE). 
Second, our survey consent form clearly states the purpose of the study, and the intended participants are RSEs. 
Third, to exclude irrelevant participants, we used exclusion criteria for those who are not involved in any research project. 
Fourth, to gather the RSEs' expertise level, we included a question (D7) that asked participants how many years they had worked as an RSE. 
While most respondents reported more than one year of experience, seven participants indicated less than one year, which may include individuals who are new to the role or with no prior experience as RSEs.

\begin{figure*}[!htb]
\caption{Survey Questions}
\label{fig_survey_questions}

\begin{tcolorbox}[enhanced,drop shadow]
\textbf{Demographics}
	\begin{description}
		\item [\hfill $\bigstar$D1] What is your gender? (Male, Female, Other/prefer not to say)
		\item [\hfill $\bigstar$D2] Do you identify as a member of any underrepresented group? If so, which one(s)? (yes, no) 
		\item [\hfill $\bigstar$D3] How frequently do you review code? (Daily, Weekly, Monthly, Yearly, Never)
		\item [\hfill $\bigstar$D4] What is the general domain of your research software project?
	    \item [\hfill $\bigstar$D5] How many FTEs (approximately) work on your project (enter 0 if unknown)?
            \item [\hfill $\bigstar$D6] Approximately how long has your project been in existence? (\textless 1 year, 1-2 years, 3-5 years, 5-10 years, \textgreater 10 years, Unknown)
		\item [D7] How many years have you worked as an RSE? (Less than 1 year, 1 years to less than 5 years, 5 years to 10 years, More than 10 years)
		\item [D8] Which roles have you performed? (write code, report bugs, fix bugs, maintain project infrastructure, make strategic decisions about direction of projects, others) 
		\item [D9] Do you receive financial compensation for your participation (e.g., as part of your job)? (Yes, No)
		\item [D10] What is the balance between “code you review” and “code you ask others to review”? (1 - Only a code reviewee, 2, 3, 4 - Equally a code reviewee and a code reviewer, 5, 6, 7 - Only a code reviewer)
	\end{description}
\vspace{8pt}
\textbf{RQ1: How do research software engineers perform peer code review?}
	\begin{description}
		\item [Q1] Which tools or techniques do you use to conduct your peer code reviews? (Pull request, Internal ticketing system, Email, Tool-based (e.g. Gerrit), Face-to-face or Pair programming, Other)
		\item [Q2] What portion of the overall code commits in the project undergo review? (Less than 10\%, 11\% - 25\%, 26\% - 50\%, 51\% - 75\%, More than 75\%)
		\item [Q3] How many people review code in a given month on average?
		\item [Q4] What factors do you consider when deciding whether to accept a peer code review request? (Leave blank if you have not reviewed code)
		\item [Q5] How many Lines of Code are typically reviewed at one time? (0 - 20 (i.e. a very small change), 21 - 50 (i.e. a small change), 51 - 100 (i.e. a medium change), 101 - 200 (i.e. a large change), \textgreater 200 (i.e. a very large change))
            \item [Q6] How many hours per week, on average, do you spend reviewing other contributors code? (Less than 1 hour, 1 - 5 hours, 6 - 10 hours, 10 - 15 hours, 16 - 20 hours, More than 20 hours)
            \item [Q7] How long, on average, does it take for someone to receive a first response after submitting a review request? (Less than 1 hour, Less than 1 day, 1-3 days, 4 - 7 days, More than 7 days)
            \item [Q8] How long, on average, does it take for someone to receive a final decision after making a review request? (Less than 1 hour, Less than 1 day, Less than 1 week, Less than 1 month, More than 1 month)
            \item [Q9] What types of problems do the code reviews identify?
            \item [\hfill $\bigstar$Q10] How often do you experience disagreements or conflicts during peer code reviews? (Always, Often, Sometimes, Rarely, Never)
            \item [\hfill $\bigstar$Q11] How important is mentorship for effective peer code reviews? (Very unimportant, Unimportant, Neither important nor unimportant, Important, Very important)
            \item [Q12] We want to know more about what makes for a good code review process. Describe any positive experiences you have had with peer code reviews.
            \item [Q13] Describe any negative experiences you have had with peer code reviews.
	\end{description}
 \end{tcolorbox}

 \end{figure*}

\begin{figure*}[!htb]
\caption{Survey Questions contd.}
\label{fig_survey_questions2}

\begin{tcolorbox}[enhanced,drop shadow]
\textbf{RQ2: What effect does peer code review have on research software?}
	\begin{description}
		\item [Q14] Do you agree or disagree with the following statement: "Peer code review is important"? (Strongly disagree, Somewhat disagree, Neither agree nor disagree, Somewhat agree, Strongly agree)
		\item [Q15] If possible, explain your answer to the previous question.
		\item [Q16] Do you agree or disagree with the following statement: "Peer code review helps improve the code"? (Strongly disagree, Somewhat disagree, Neither agree nor disagree, Somewhat agree, Strongly agree)
		\item [Q17] If possible, explain your answer to the previous question.
		\item [Q18] Do you agree or disagree with the following statement: "Peer code review helps decrease code complexity"? (Strongly disagree, Somewhat disagree, Neither agree nor disagree, Somewhat agree, Strongly agree)
		\item [Q19] If possible, explain your answer to the previous question.
            \item [\hfill $\bigstar$Q20] Do you agree or disagree with the following statement: “Peer code review helps find the performance bottlenecks or optimization opportunities”? (Strongly disagree, Somewhat disagree, Neither agree nor disagree, Somewhat agree, Strongly agree)
            \item [Q21]	If possible, explain your answer to the previous question.
            \item [\hfill $\bigstar$Q22] Do you agree or disagree with the following statement: “Peer code review produces more bug-free research software”? (Strongly disagree, Somewhat disagree, Neither agree nor disagree, Somewhat agree, Strongly agree)
            \item [Q23] If possible, explain your answer to the previous question.
            \item [\hfill $\bigstar$Q24] Do you agree or disagree with the following statement: “Peer code review increases the overall maintainability of the research software”? (Strongly disagree, Somewhat disagree, Neither agree nor disagree, Somewhat agree, Strongly agree)
            \item [Q25] If possible, explain your answer to the previous question.
            \item [\hfill $\bigstar$Q26] Do you agree or disagree with the following statement: “Peer code review has a positive impact on the reduction of technical debt”?	(Strongly disagree, Somewhat disagree, Neither agree nor disagree, Somewhat agree, Strongly agree)
            \item [Q27] If possible, explain your answer to the previous question.
	\end{description}
 \vspace{8pt}
\textbf{RQ3: What difficulties do research software engineers face with peer code review?}
	\begin{description}
		\item [Q28] What aspects of the peer code review process are challenging or what barriers do reviewers face when reviewing code?
	\end{description}
\vspace{8pt}
\textbf{RQ4: What improvements to the peer code review process do research software engineers need?
}
	\begin{description}
	    \item [Q29] Is there anything missing from the peer code review process?
		\item [Q30] How could the peer code review process be improved?
	\end{description}
\end{tcolorbox} \end{figure*}

\subsection{Data Analysis}
Before performing the data analysis, the first author reviewed the complete survey responses to understand the data. 
Similar to our previous work: a valid response answered all the quantitative questions and at least one qualitative question. 
The first author prepared a list of the 61 valid responses.

We used Python to generate frequency distributions for the quantitative answers. 
For the qualitative data, we prepared a Google spreadsheet and followed a standard qualitative analysis approach to code the responses. 
First, the first and second authors individually mapped the responses to one or more codes. 
Then, the first author identified any dissimilar codes, which the first and second authors discussed to resolve the disagreement.
If they were unable to resolve it, the third author provided input to help resolve it.

There were two types of disagreements. 
First, we had \textit{semantic disagreements}, in which the two authors interpreted the meaning of the answer differently.
Second, we had \textit{syntacic disagreements}, in which the authors used different words for the same underlying concept.
In both cases, we resolved all disagreements in the final deataset.
Once our coding was complete, we used Python to calculate the frequency of each unique code and visualize the distribution in charts.

\section{Results}
\label{sec:Results}
In this section, we present the results organized around our four research questions.
For each, we draw insights from the current RSE survey and compare those results with the previous survey results, where applicable. 
In addition to the questions from the previous survey, we introduced new questions. 
These questions help us better understand our participants' expertise, how they perform code review, and their perception of the effect of code review on research software.
Because the number of respondents differed between the current study and the previous study, to make it easier to compare the results, we converted the raw numbers into percentages based on the number of valid responses in each survey.
When analyzing free responses, we applied multiple codes to individual answers, so response counts often exceed 100\%. 
For some questions, we received both positive and negative reactions.
In those cases, the respondents' explanations may have included similar reasons, resulting in the same codes appearing in both the positive and the negative explanations. 

The survey question numbers in this section correspond to those in Figures~\ref{fig_survey_questions} and \ref{fig_survey_questions2}.

\subsection{\textit{\textbf{Demographics}}}
The goal of this section is to characterize the sample of survey respondents.
Where possible, we compare the sample with that from the previous survey.
Overall, these demographics show that the respondents have sufficient experience and engagement with the peer code review process to provide informed feedback on its use in their projects.

Regarding gender, the majority of respondents were male (77\% male vs. 20\% female).
We also had 16\% from underrepresented groups. 

\begin{figure*}[!htb]
	\includegraphics[width=1.0\textwidth]{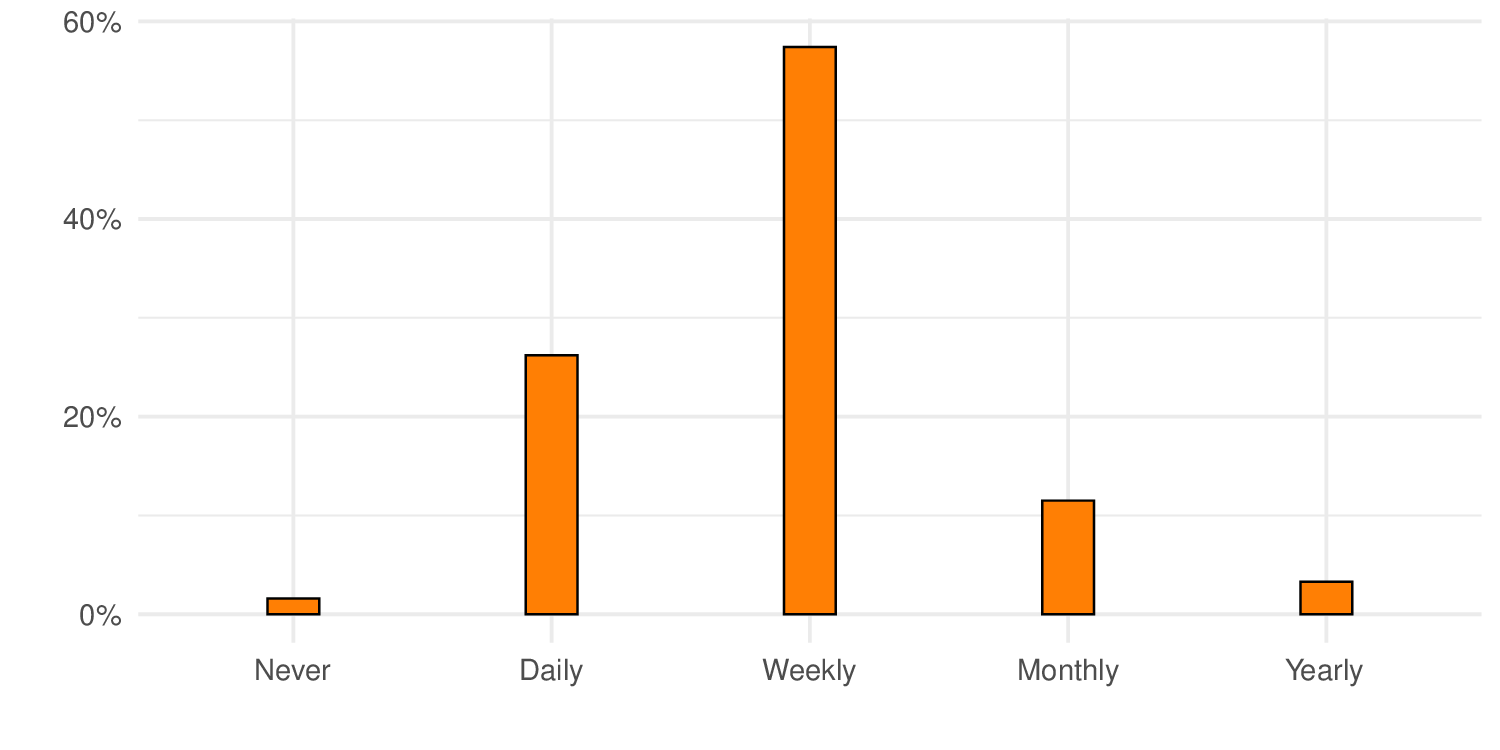}
	\caption{Frequency of code reviews by RSEs [D3]}
	\label{fig:D3}
\end{figure*}

As shown in Figure~\ref{fig:D3}, 84\% of respondents report conducting code reviews at least once a week, indicating they have the experience to provide valuable insights.

We identified RSEs from 55 domains, which we grouped into the higher-level categories in Table~\ref{tab:domain}.
The diversity of domains helps the results be more generalizable.

\begin{figure*}[!htb]
	\includegraphics[width=1.0\textwidth]{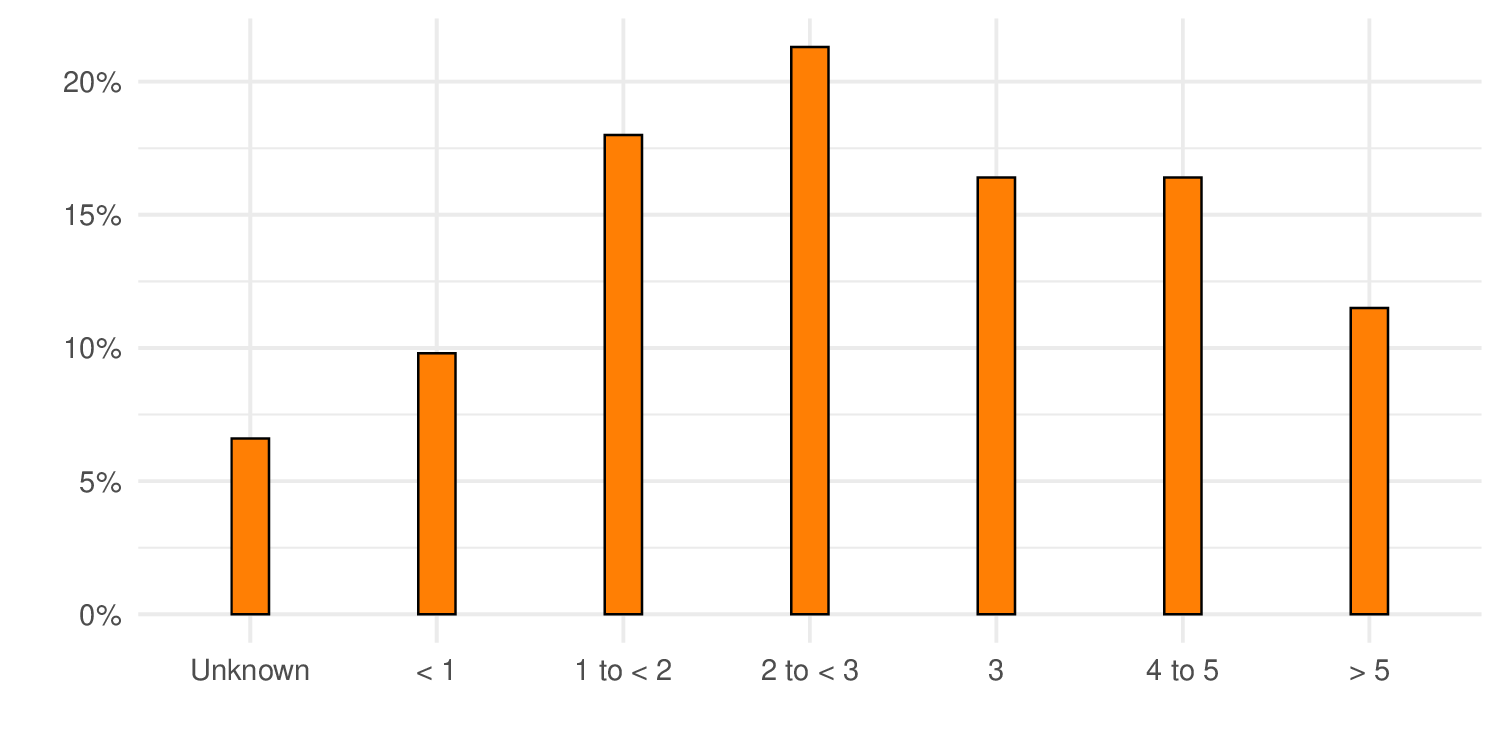}
	\caption{Number of FTEs working on research software projects [D5]}
	\label{fig:D5}
\end{figure*}

\begin{longtable}{|p{.2\textwidth}|c|p{0.6\textwidth}|}
\caption{Domains of RSE Projects [D4]} \label{tab:domain} \\ 
\hline
\textbf{Category} & \textbf{Count} & \textbf{Domains} \\ 
\hline
\endfirsthead

\hline
\textbf{Category} & \textbf{Count} & \textbf{Domains} \\ 
\hline
\endhead

\hline
\endfoot

Health Science & 13 & Biomedical Engineering, Infectious Disease Epidemiology, Epidemiology / R Development, Health Research, Clinical Trials, Medical Research / Data Extraction, Bioinformatics / Metabolomics, Bioinformatics, Biomedical Databases, Bioimaging, Neuroscience, Electrophysiology Data Analysis \\ \hline
Physics & 12 & Astronomy, Astrophysics, Astronomy / Astrophysics, Heliophysics, Physics / Data Analysis and Modeling, Physics, Geophysics, Earth System Science \\ \hline
Social Science	& 9	& Psychology, Behavioral Science, Digital Phenotyping, Citizen Science, Educational Technology (EdTech), Science Communication / Human-Computer Interaction (HCI), Library and Information Science, Digital Humanities \\ \hline
Computational Science & 7 & Distributed Computing, Numerical Linear Algebra / High-Performance Computing, High-Performance Computing, GPU-Accelerated Simulation, Neuromorphic Computing, Bayesian Statistics, Numerical Linear Algebra \\ \hline
Computer Science & 6 & Workflow Management / Software Engineering, Computer Vision, Computer Vision / GUI Development, Web Development, Infrastructure \\ \hline
Chemistry & 4 & Chemical Engineering / Mathematics, Chemical Engineering, Molecular Simulation, Molecular Simulations \\ \hline

Environmental Science & 4 & Weather and Climate Modeling, Atmospheric / Climate Science Data Analysis, Remote Sensing, Climate Change Research \\ \hline
Data Science & 4 & Metadata Management, Spatio-Temporal Data Analysis / Data Visualization, Data Science, Machine Learning / Cryogenic Macroscopy \\ \hline

Others & 3 & Process Systems Engineering, Finite Element Method \\ \hline
\end{longtable} 
As Figure~\ref{fig:D5} shows, most of the RSE's projects are relatively small, with between 1 and 5 FTEs.

The majority of the RSEs (72\%) work on projects that are older than 3 years, suggesting the projects have lasted long enough to make use of peer code review.

\begin{figure*}[!htb]
	\includegraphics[width=1.0\textwidth]{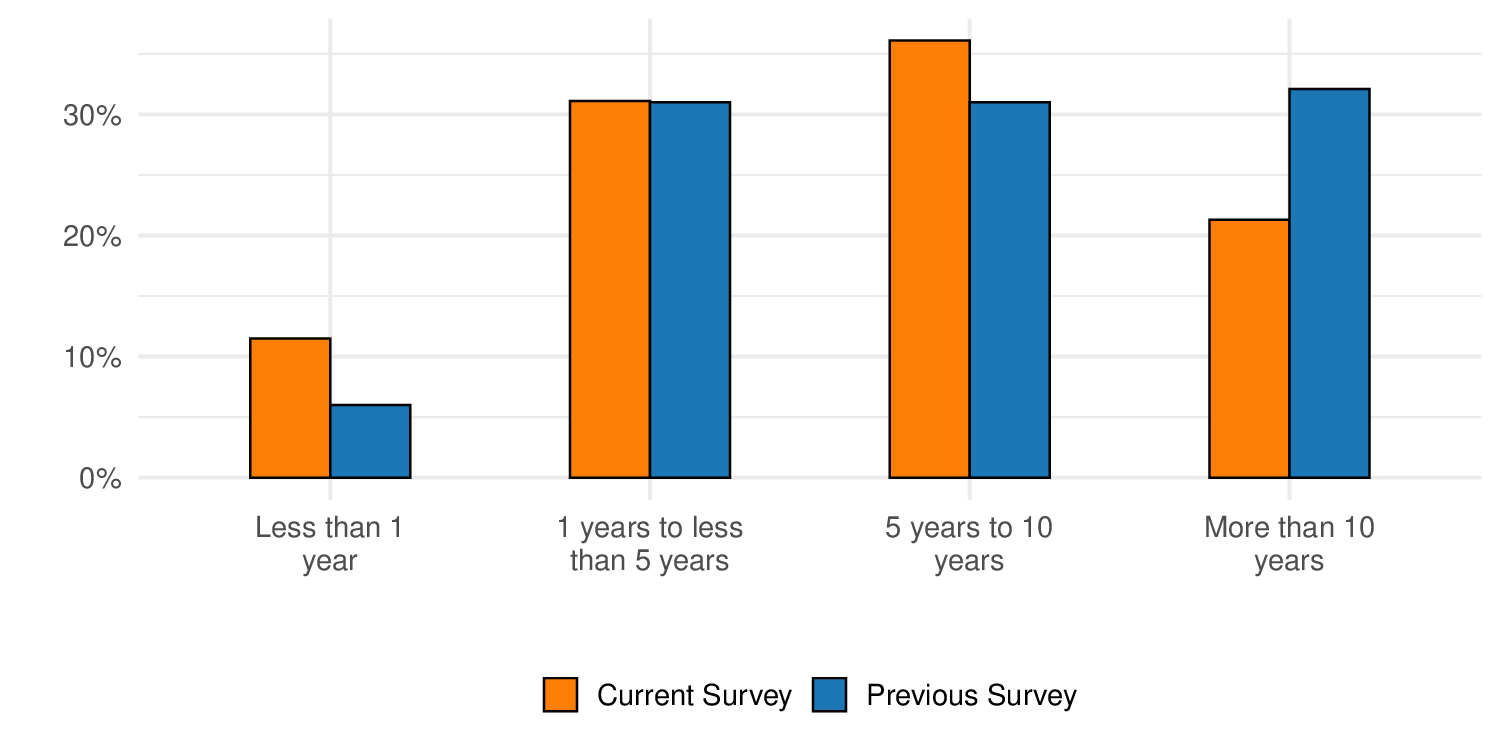}
	\caption{Number of years worked as an RSE vs. developers [D7]}
	\label{fig:D7}
\end{figure*}

The results in Figure~\ref{fig:D7} show a majority of RSEs have been working on their research projects for over a year.
This distribution is similar to that from the previous survey.

Table~\ref{tab:roles_comparison} shows the various roles respondents play on their projects.
This distribution is similar to the previous survey.

\begin{table}[!htb]
\centering
\caption{Roles performed by RSEs and developers in their projects [D8]}
\label{tab:roles_comparison}
\begin{tabular}{|l|c|c|}
\hline
\textbf{Roles} & \textbf{RSE (\%)} & \textbf{Developer (\%)} \\ \hline
Add new code & 100 & 89.3 \\ \hline
Fix bugs & 96.7 & 79.8 \\ \hline
Report bugs & 88.5 & 73.8 \\ \hline
Maintain infrastructure & 86.9 & 67.9 \\ \hline
Make decisions & 86.9 & 67.9 \\ \hline
Other & 57.4 & 20.2 \\ \hline
\end{tabular} \end{table}

In response to D9, 97\% of the respondents received financial compensation for their work, while only 70\% of the previous respondents did.

\begin{figure*}[!htb]
	\includegraphics[width=1.0\textwidth]{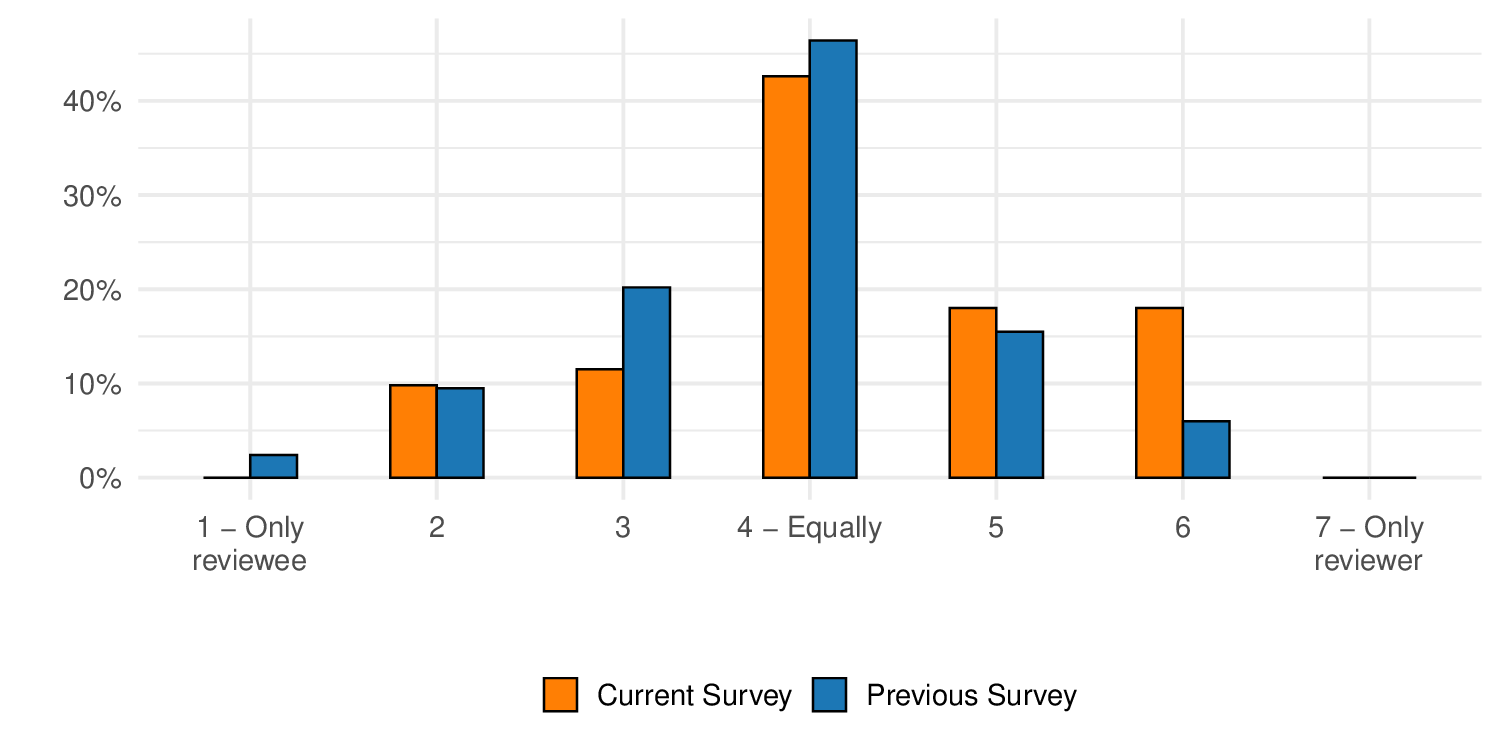}
	\caption{Balance as a reviewer and reviewee: RSEs vs. developers [D10]}
	\label{fig:D10}
\end{figure*}

As Figure~\ref{fig:D10} shows, respondents participate as both reviewers and reviewers. 
This distribution is similar to the distribution from the previous survey, with a slight shift to the \textit{reviewer} side.

\subsection{RQ1: How do research software engineers perform peer code review?}
In this section, rather than report the results of each question separately, we group related questions around several themes.

\subsubsection{Theme 1: Overall practices}
The majority of RSEs (59\%) conduct code reviews via pull requests on platforms like GitHub or GitLab [Q1].
This result is similar to the previous survey (54\%).

\begin{figure*}[!htb]
	\includegraphics[width=1.0\textwidth]{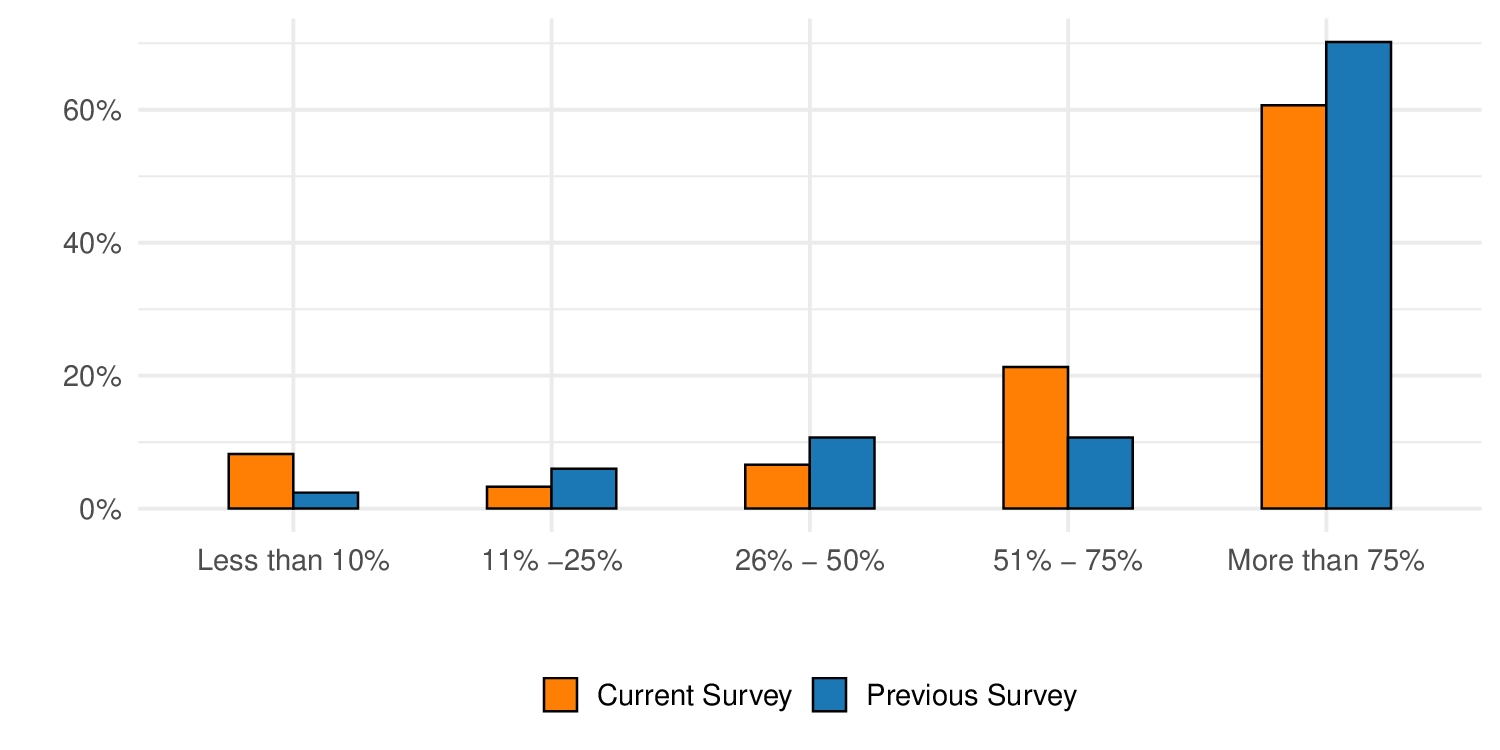}
	\caption{Portion of code reviewed per month by RSEs vs. developers [Q2]}
	\label{fig:Q2}
\end{figure*}

As Figure~\ref{fig:Q2} shows, a large percentage of the code undergoes peer code review.
This distribution is similar to the results from the first survey.

\begin{figure*}[!htb]
	\includegraphics[width=1.0\textwidth]{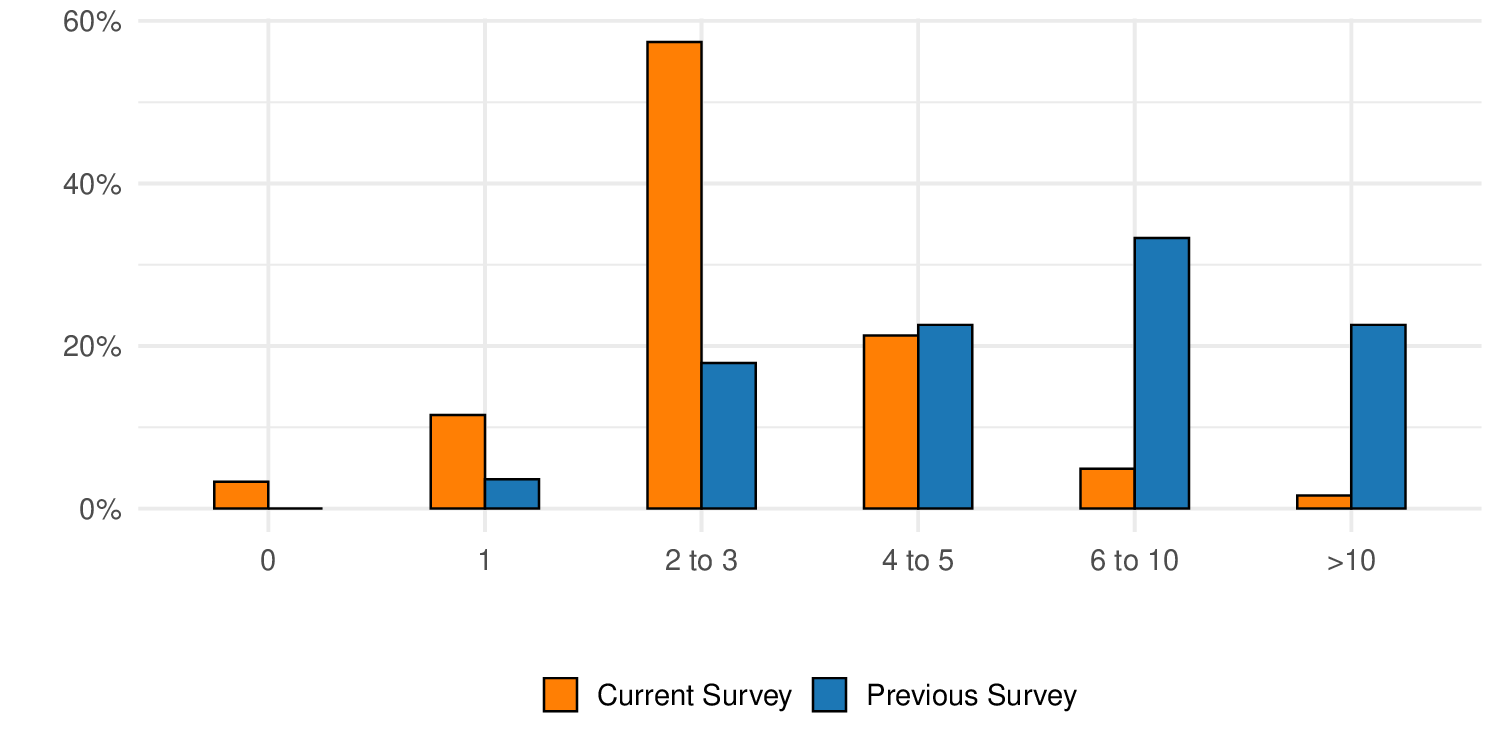}
	\caption{Average number of reviewer per month: RSEs vs. Developers [Q3]}
	\label{fig:Q3}
\end{figure*}

Figure~\ref{fig:Q3} shows the majority of RSEs (around 70\%) observe that code reviews are typically conducted by 1 to 3 people.
In the previous survey, more team members reviewed code.

\subsubsection{Theme 2: Acceptance of review requests}
\begin{figure*}[!htb]
	\includegraphics[width=1.0\textwidth]{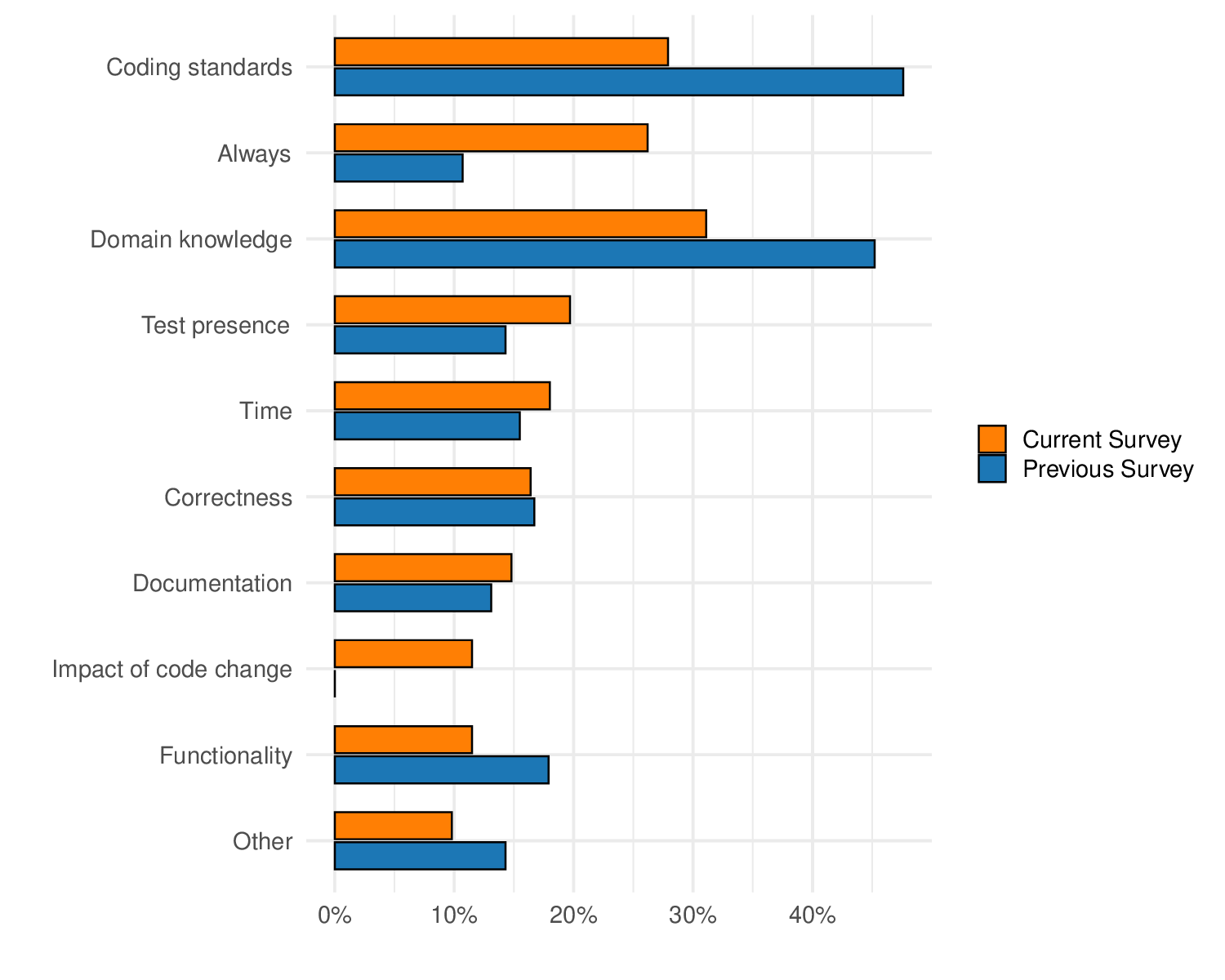}
	\caption{Factors for accepting peer code review request: RSEs vs. developers [Q4]}
	\label{fig:Q4}
\end{figure*}

Figure~\ref{fig:Q4} shows the distribution of responses to Q4 (coded from free-responses), which asked respondents to describe factors influencing whether they accept a review request.
There are a number of factors with very similar percentages of responses, suggesting there are many reasons why RSEs accept requests.
One note: the \textit{always} code captures responses that indicate review requests were always accepted.

There are some notable differences between this survey and the previous one.
In the previous survey, \textit{coding standards} and \textit{domain knowledge} were more prevalent.
It is likely that for RSEs, more code follows coding standards, and they review code for projects with which they are familiar.
In the current survey, we saw one reason that was not present in the first survey: \textit{impact of the code change}, with one respondent explaining, ``whether code is in a public or private API, the estimated number of downstream users that will be affected, and the severity of effect or impact of possible bugs and behavior changes in the code path being changed.''

\subsubsection{Theme 3: Amount of Code Reviewed}
\begin{figure*}[!htb]
	\includegraphics[width=1.0\textwidth]{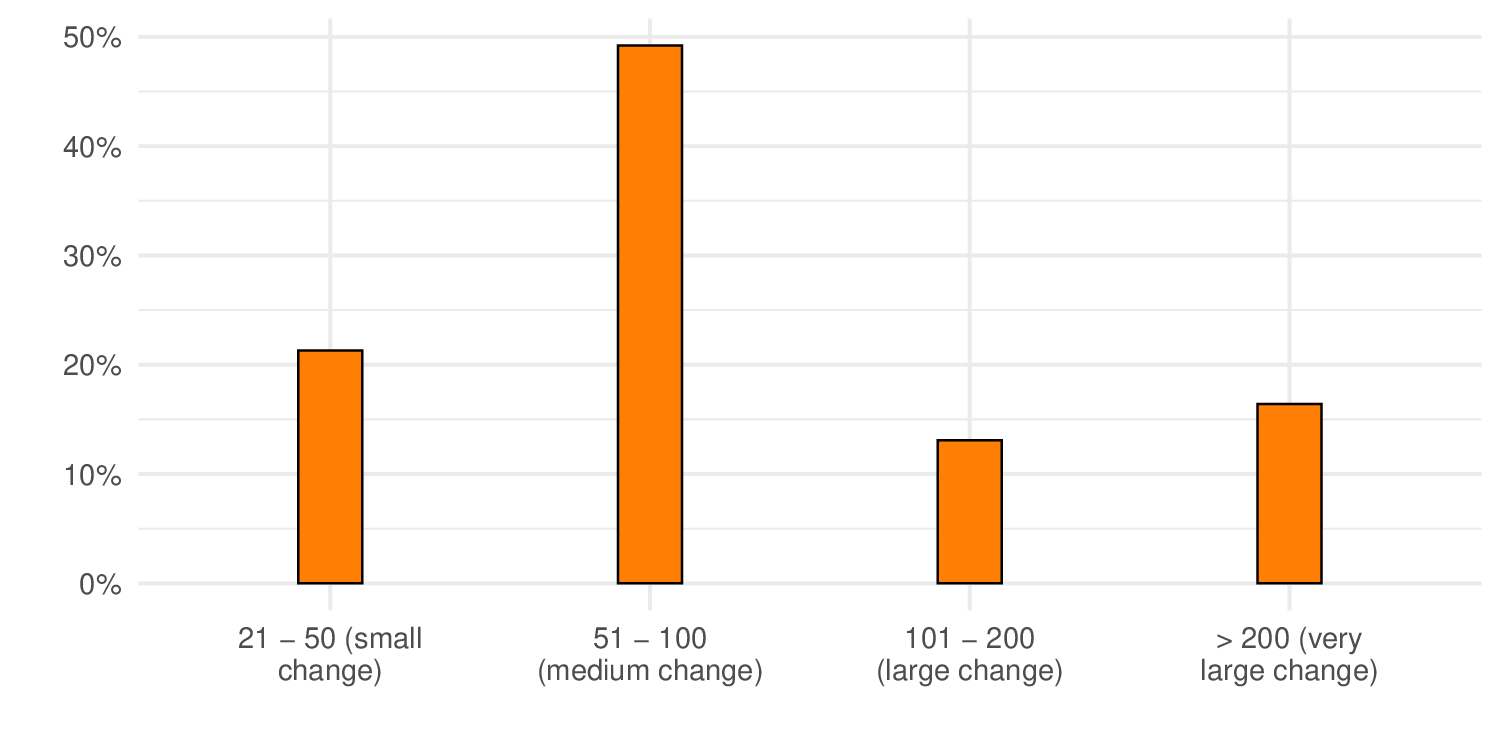}
	\caption{Average lines of code reviewed at one time by RSEs [Q5]}
	\label{fig:Q5}
\end{figure*}

As Figure~\ref{fig:Q5} shows, most people (71\%) review small to medium-sized changes.
However, surprisingly, 29\% review large and very large changes.

\subsubsection{Theme 4: Time Spent Reviewing Code}
\begin{figure*}[!htb]
	\includegraphics[width=1.0\textwidth]{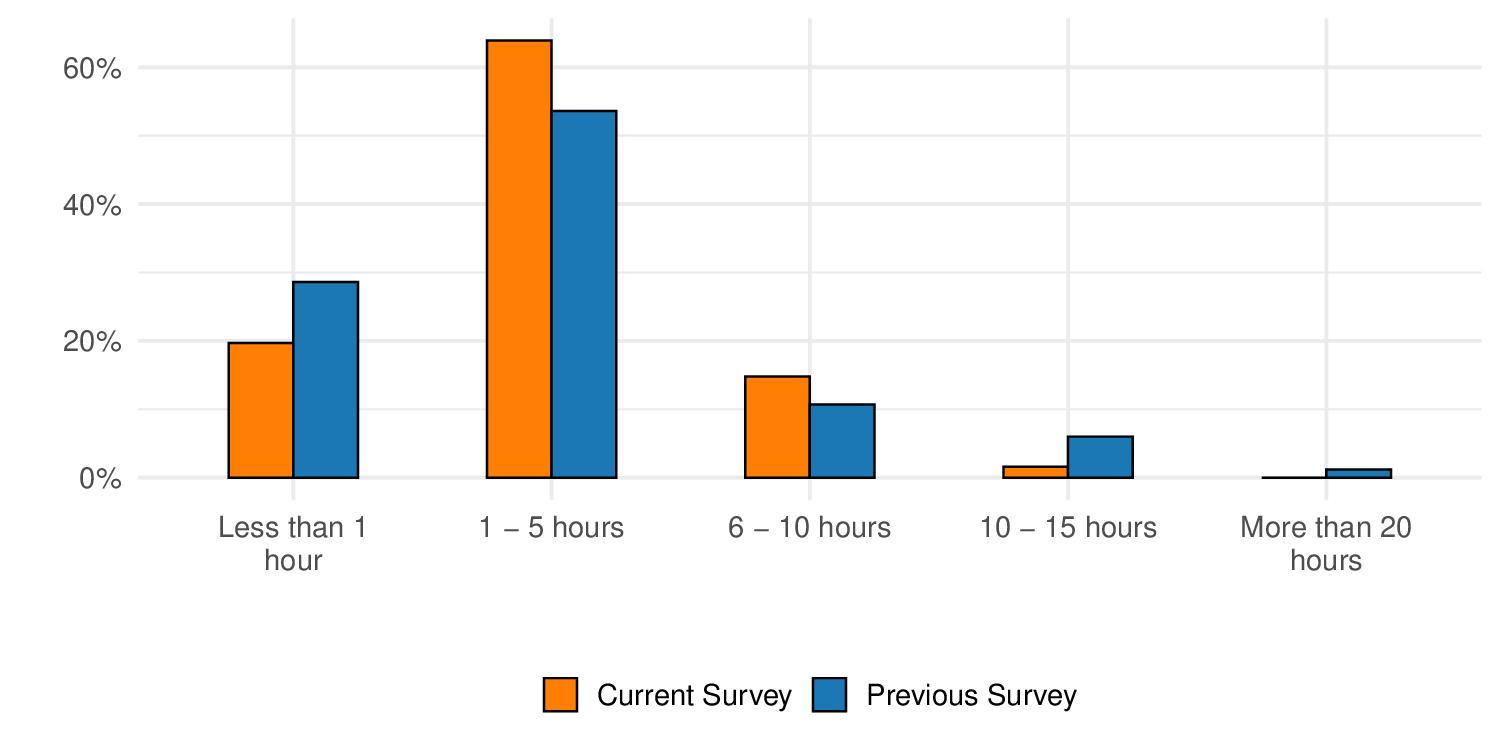}
	\caption{Time spent on peer code review: RSEs vs. developers [Q6]}
	\label{fig:Q6}
\end{figure*}

Figure~\ref{fig:Q6} illustrates the amount of time reviewers spend on code reviews. 
Most respondents (84\%) spent up to 5 hours per week on code review.
This result is similar to the results from the previous survey.

\begin{figure*}[!htb]
	\includegraphics[width=1.0\textwidth]{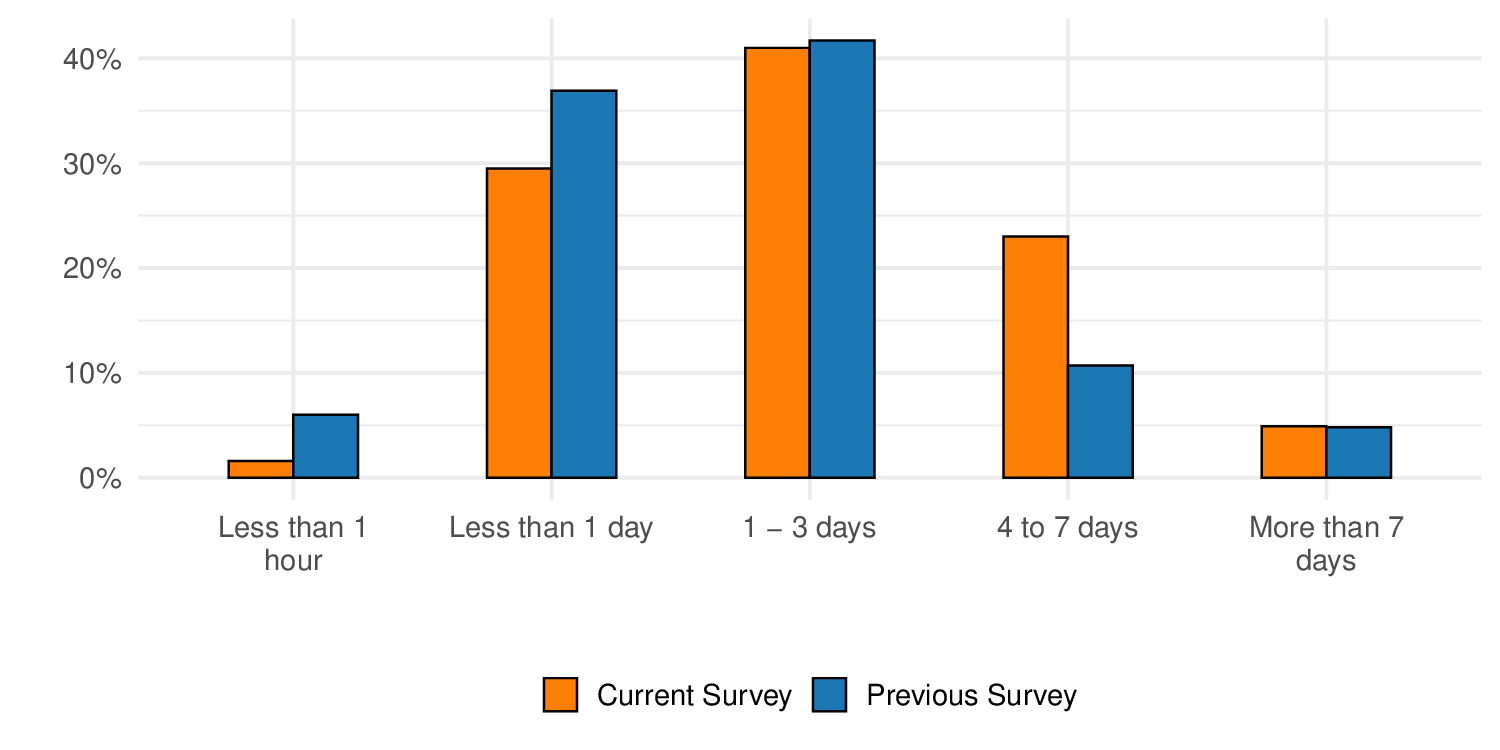}
	\caption{Average time to receive a first response: RSEs vs. developers [Q7]}
	\label{fig:Q7}
\end{figure*}

Figure~\ref{fig:Q7} shows how long authors have to wait before receiving a response to their review request. 
The vast majority of respondents receive a response in less than a week.
Compared with the previous survey, RSEs may wait a bit longer on average, but approximately the same small percentage wait longer than week.

\begin{figure*}[!htb]
	\includegraphics[width=1.0\textwidth]{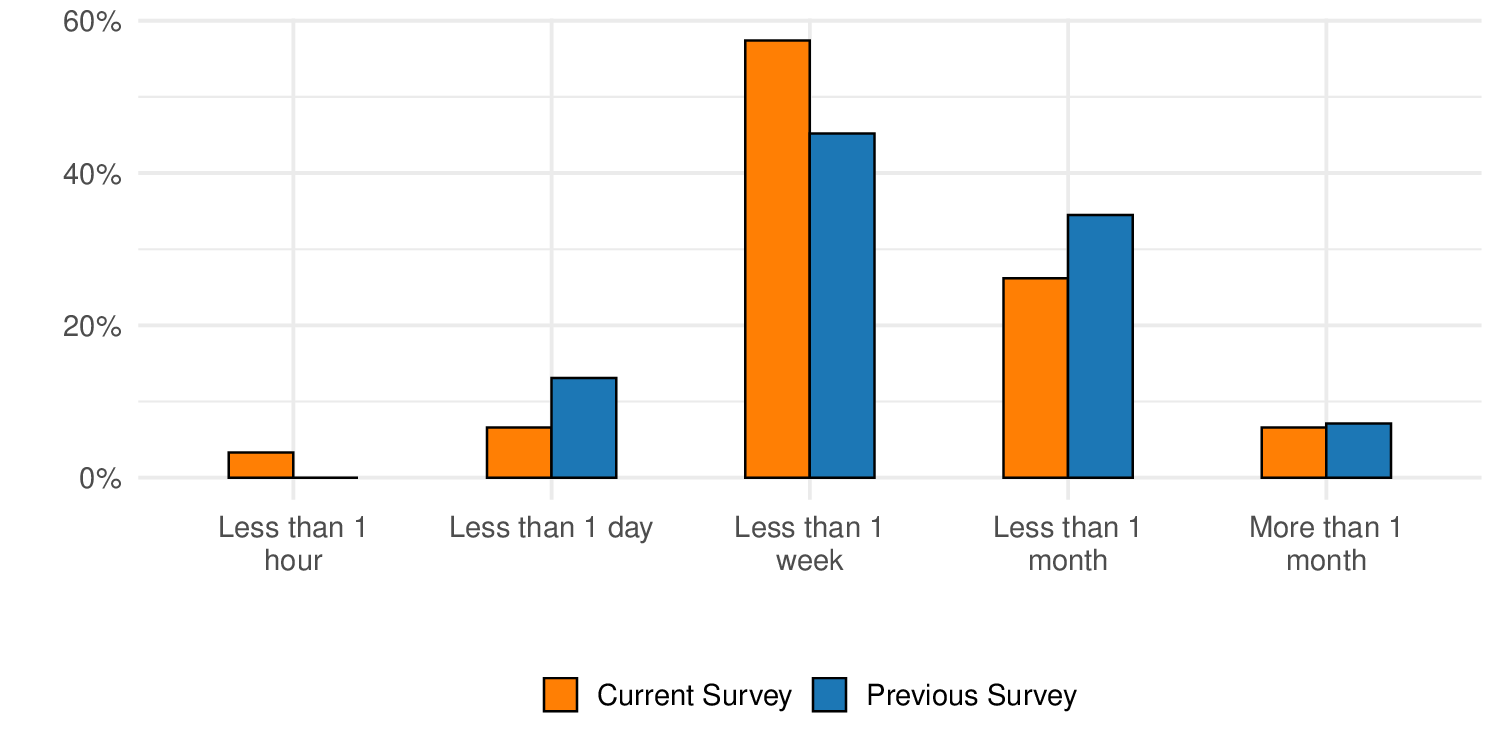}
	\caption{Average time to receive a final decision: RSEs vs. developers [Q8]}
	\label{fig:Q8}
\end{figure*}

However, even if many have to wait 4-7 days for an initial response, as Figure~\ref{fig:Q8} shows, most RSEs receive a final decision in less than a week.
Compared with the results from the previous survey, RSEs receive a final decision quicker, on average.

\subsubsection{Theme 5: Problems Identified During Code Review}
\begin{figure*}[!htb]
	\includegraphics[width=1.0\textwidth]{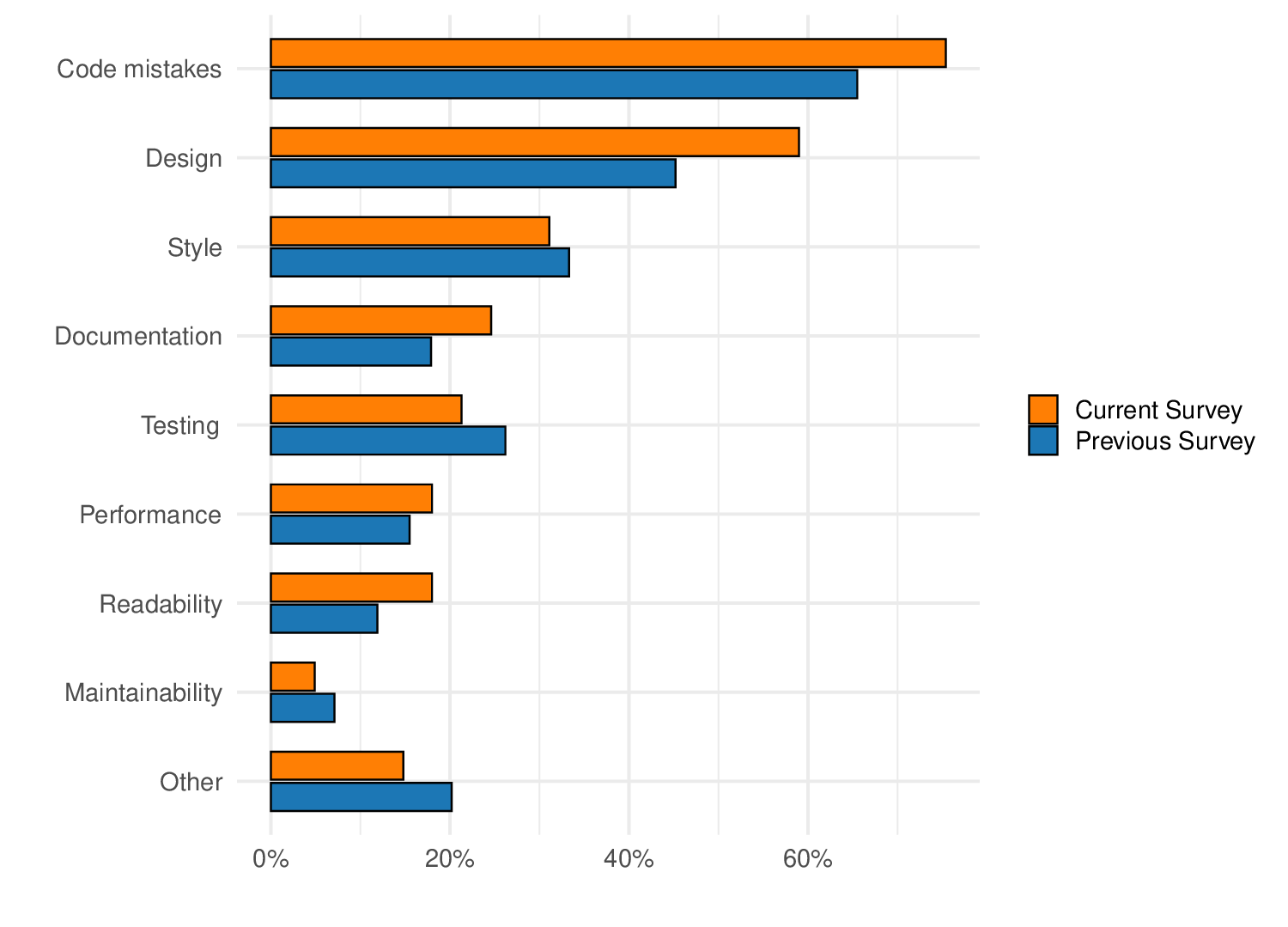}
	\caption{Problems peer code reviews identify: RSEs vs developers [Q9]}
	\label{fig:Q9}
\end{figure*}

As Figure~\ref{fig:Q9} shows, the most common types of problems identified are \textit{code mistakes} and \textit{design} issues, suggesting that the primary focus of the reviewers is on correctness. 
The distribution of responses is similar to those from the previous survey.

\subsubsection{Theme 6: Interactions With Teammates} 
\begin{figure*}[!htb]
	\includegraphics[width=1.0\textwidth]{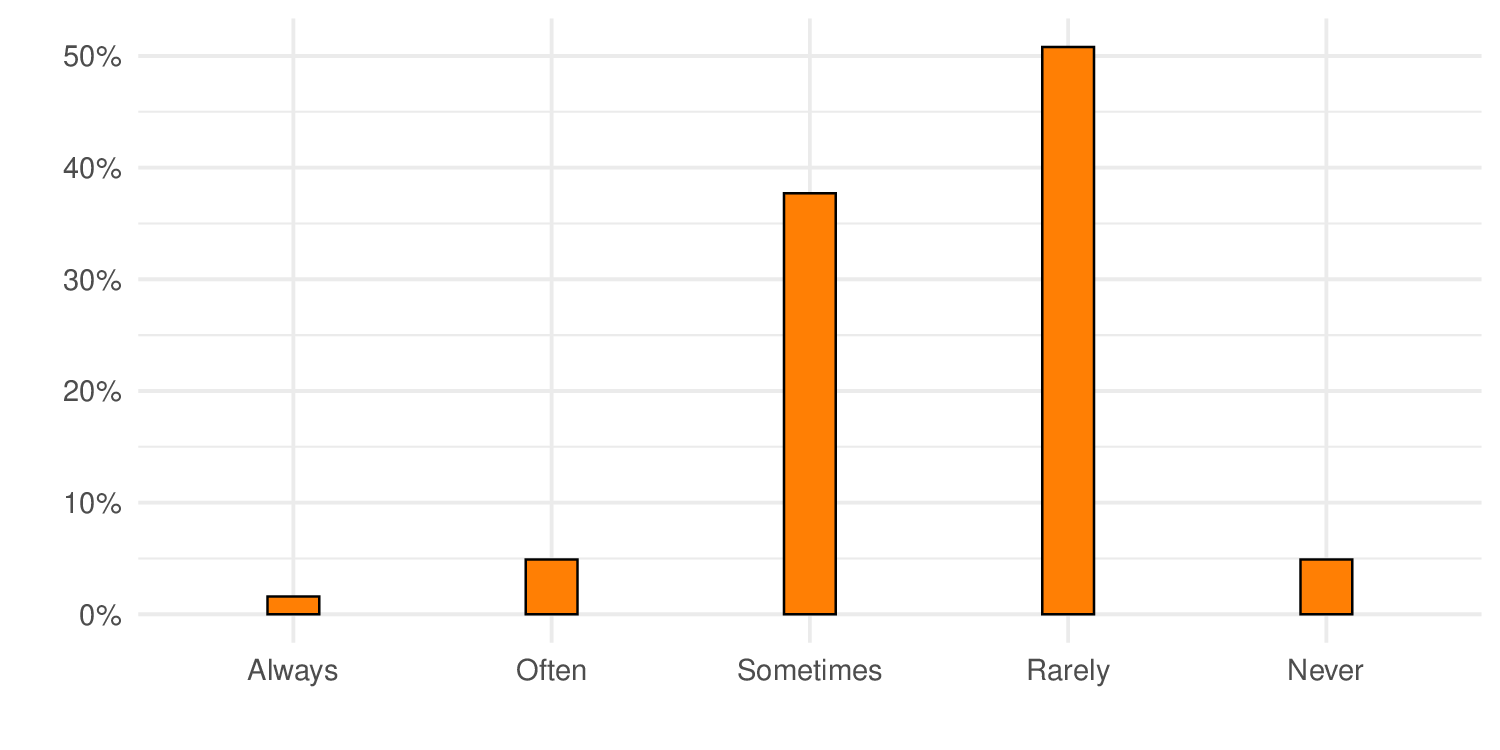}
	\caption{Frequency of disagreements during peer code reviews by RSEs [Q10]}
	\label{fig:Q10}
\end{figure*}

Figure~\ref{fig:Q10} shows the frequency of disagreements is relatively low, with 56\% responding \textit{rarely} or \textit{never}.
In addition, 78\% of the respondents found mentoring in peer code review to be \textit{important} or \textit{very important}.
These results suggest that the generally positive interactions with teammates are important to the success of peer code review.

\subsubsection{Theme 7: Positive peer code review experiences}
\begin{figure*}[!htb]
	\includegraphics[width=\textwidth]{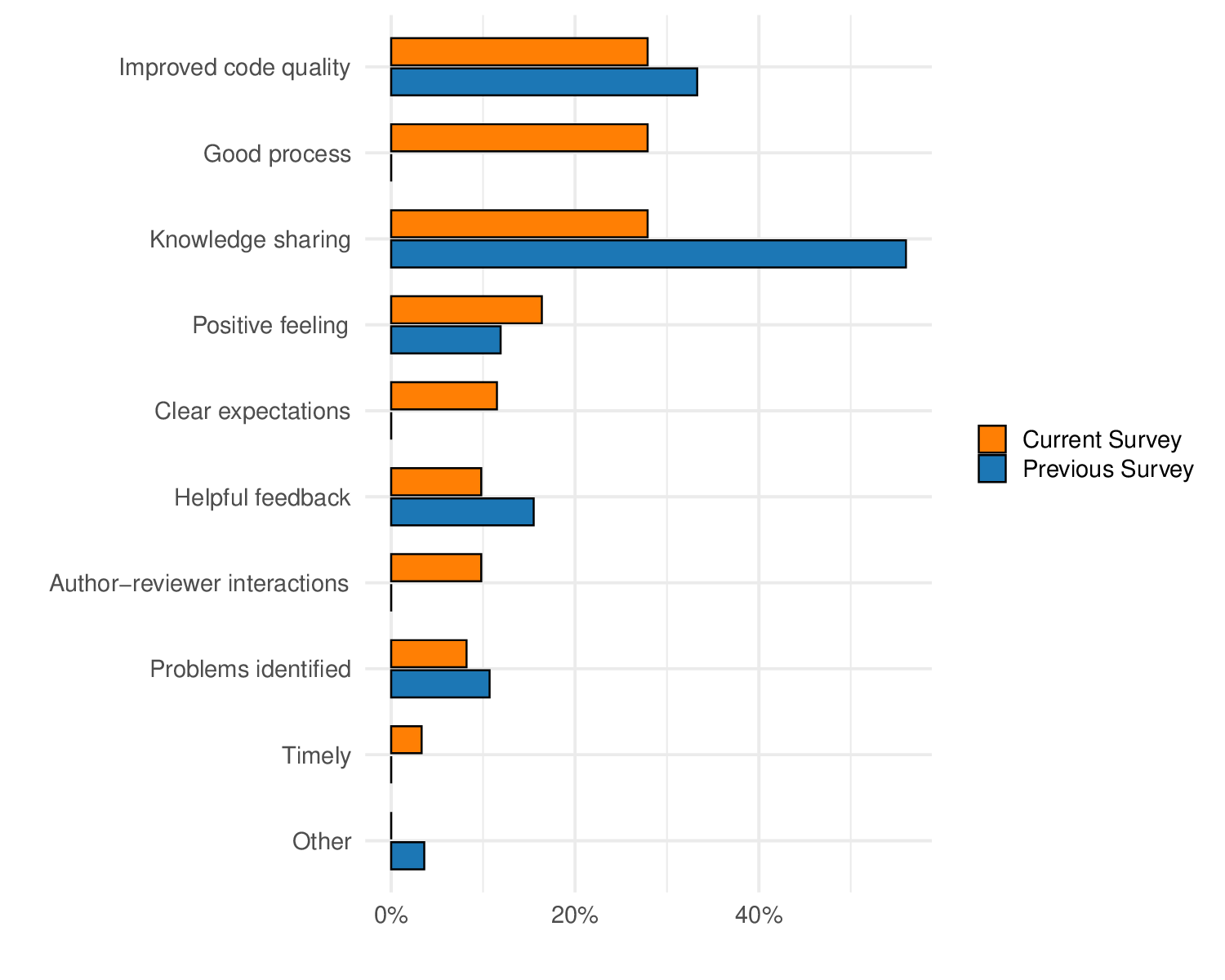}
	\caption{Positive experiences with peer code reviews: RSEs vs developers [Q12]}
	\label{fig:Q12}
\end{figure*}

Figure~\ref{fig:Q12} shows the coded results from the free-response question about positive experiences.
The RSEs valued \textit{improved code quality}, a \textit{good process}, and \textit{knowledge sharing}. 
As one RSE explained, ``If the code review process extends over a longer period, I've observed that the students submitting their work learn from it and consequently write better code.'' 
Another noted code review being a good process as, ``We are common sense driven. We review the code imagining that we have to maintain this code in the future. Following common sense, a set of conventional practices will be established automatically.''
This result is consistent with the others that showed value in both technical and social dimensions.
The most glaring difference with the previous results was the much higher frequency of responses about \textit{knowledge sharing} from the previous survey.

\subsubsection{Theme 8: Negative peer code review experiences}
\begin{figure*}[!htb]
\centering
	\includegraphics[width=1.0\textwidth]{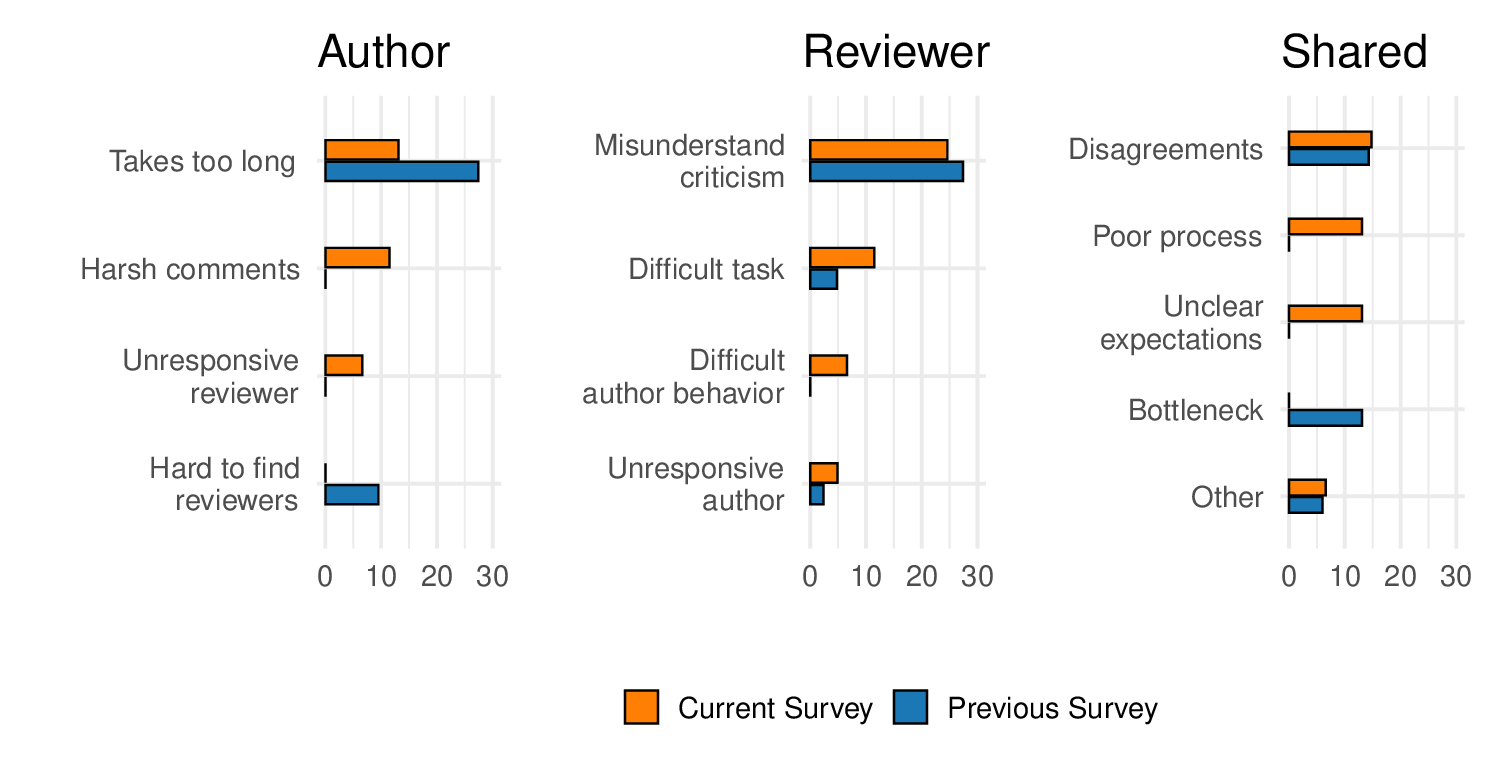}
	\caption{Negative experiences with peer code reviews: RSEs vs developers [Q13]}
	\label{fig:Q13}
\end{figure*}

Figure~\ref{fig:Q13} shows the coded results from the free-response question about negative experiences.
There was not one dominant negative experience.
Multiple responses appeared with relatively similar frequency. 
We grouped the responses into three perspectives: author, reviewer, and both. 
From the author's perspective, one respondent pointed out the issues with \textit{harsh comments} and \textit{unresponsive reviewer} that, ``This was not on my current project; but in some cases I've experienced highly antagonistic code reviews. I've also had many experiences where a reviewer simply stopped engaging.''
From the reviewer's perspective, we found that \textit{misunderstanding criticism} is the most common issue. 
As one respondent noted, ``Comments on the PRs can be rather dry and can be perceived as negative, even if most of the time that's not the intention of the reviewers.'' 
Another issue is related to \textit{author behavior}, as respondents notice that negative traits of authors, such as inflexibility, making repeated mistakes, or resisting change, further detract from the review experience.

Both author and reviewer experience \textit{disagreements} in their code review process with one respondent saying, ``Sometimes disagreements lead to a converged solution, and sometimes a negotiated compromise. In the worst case however, an impasse is reached, in which case either seniority takes precedence or a democratic vote is taken.''
They perceive the process as flawed due to inadequate systems, poor communication, and poor Git hygiene. 
Additionally, they face unclear expectations that arise from misalignment or a lack of clarity regarding the review's purpose. A significant concern for RSEs involves rude reviewers, whose harsh comments can feel personally attacking and negatively impact their experience. 

The biggest differences with the previous results are related to time (i.e., \textit{takes too long}, \textit{bottleneck}, and \textit{hard to find reviewers}).
Both of these responses make sense given the generally lower software engineering expertise of the previous respondents.

\subsection{RQ2: What effect does peer code review have on research software?}
Like the previous section, we group related questions around several themes.

\subsubsection{Theme 1: Peer code review is important}
\begin{figure*}[!htb]
	\includegraphics[width=1.0\textwidth]{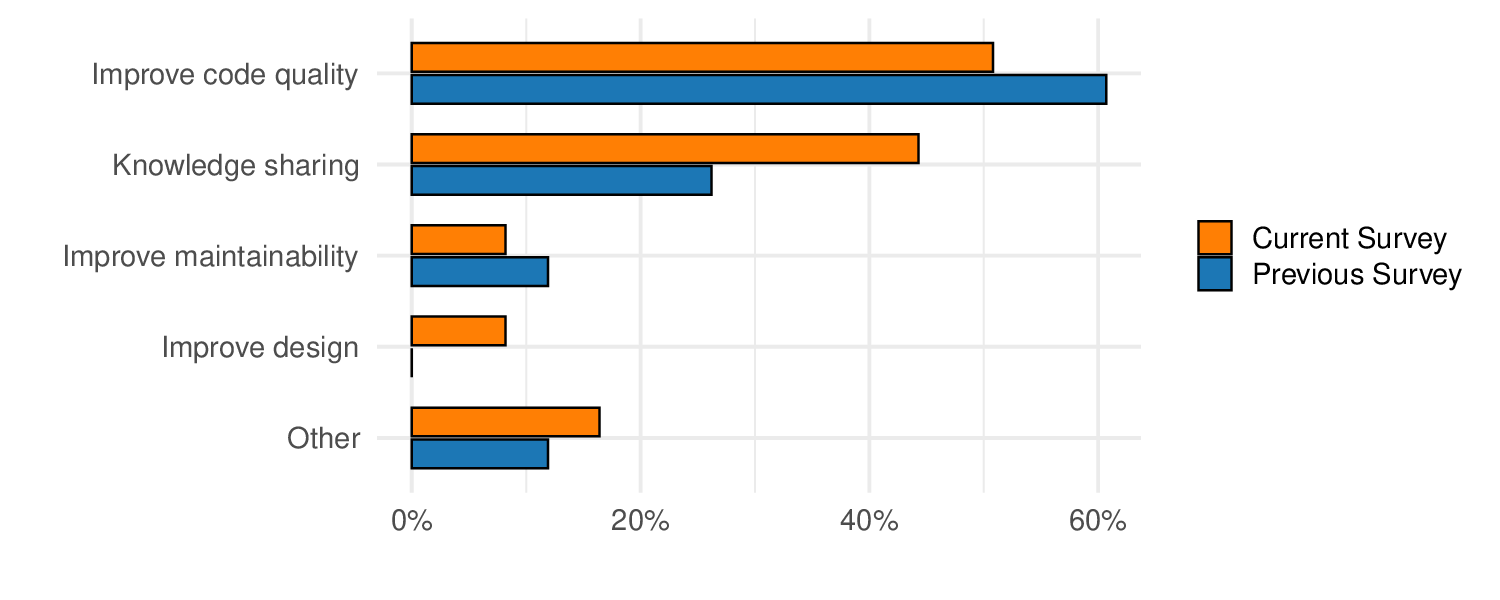}
	\caption{Reasons why peer code review is important: RSEs vs. developers [Q15]}
	\label{fig:Q15}
\end{figure*}

The vast majority of respondents (85\%) \textit{strongly agreed} that peer code review was important.
This result matches the result from the previous study where 88.1\% also \textit{strongly agreed}.
Figure~\ref{fig:Q15} shows the results of the free-response question where respondents could explain why peer code review was important.
The two most common responses are \textit{improve code quality} and \textit{knowledge sharing}, which match the prior survey results. 
As one respondent explained, ``Code written in isolation is at higher risk of failure and unsustainability. Reviews share knowledge and improve quality.'' 
Others emphasized additional benefits such as \textit{improved maintainability}: ``Having a second set of eyes look over the code helps enhance maintainability'' and the discussion of design results in \textit{improved design}: ``My code has gotten a lot better after code reviews, both in finding bugs but also in knowing my design decisions make sense to others''.

\subsubsection{Theme 2: Peer code review improves code}
\begin{figure*}[!htb]
	\includegraphics[width=1.0\textwidth]{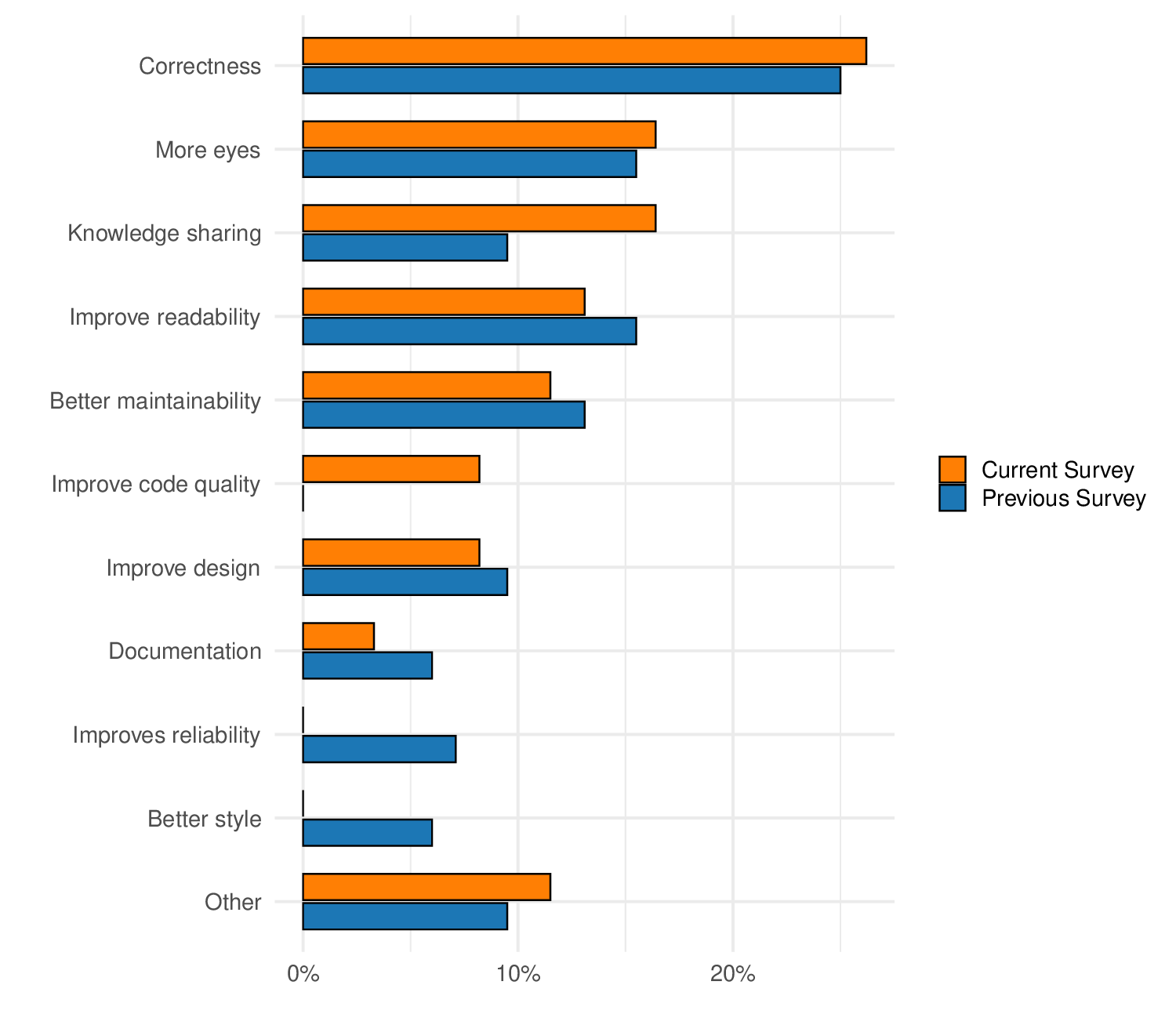}
	\caption{RSEs vs. developers' explanations of how peer code review helps improve code [Q17]}
	\label{fig:Q17}
\end{figure*}

Almost all of the respondents (99\%) \textit{strongly agree} or \textit{somewhat agree} that peer code review improves code.
This result is consistent with the results of the previous survey.
Figure~\ref{fig:Q17} shows the results of the free-response question where respondents could explain how peer code review improves code.
The most common response was that it improved \textit{correctness}. 
As one respondent noted, ``We all have blind spots or areas of less knowledge, and reviewing from others allows us to improve our own experience and correct specific issues.'' 
Respondents also explained that code review improves code in terms of overall quality, with one noting, ``Explicit discussion on design choices as well as core clarity usually result in better code.''
In addition, many responses were given by similar numbers of respondents, suggesting there are multiple ways in which peer code review improves code.
These responses included \textit{better maintainability}: ``Having a second set of eyes look over the code helps enhance maintainability'') and \textit{improved readability}: ``You cannot test readability alone… You must have another person act as tester'').
Some respondents mentioned that code review helps \textit{improve code quality} because they think, ``not everyone on the team has the same understanding of quality. It's important, but not everyone knows how to enforce it.''
Overall, the results from this survey are similar to those from the previous survey.

\subsubsection{Theme 3: Peer code review decreases code complexity}
\begin{figure*}[!htb]
	\includegraphics[width=1.0\textwidth]{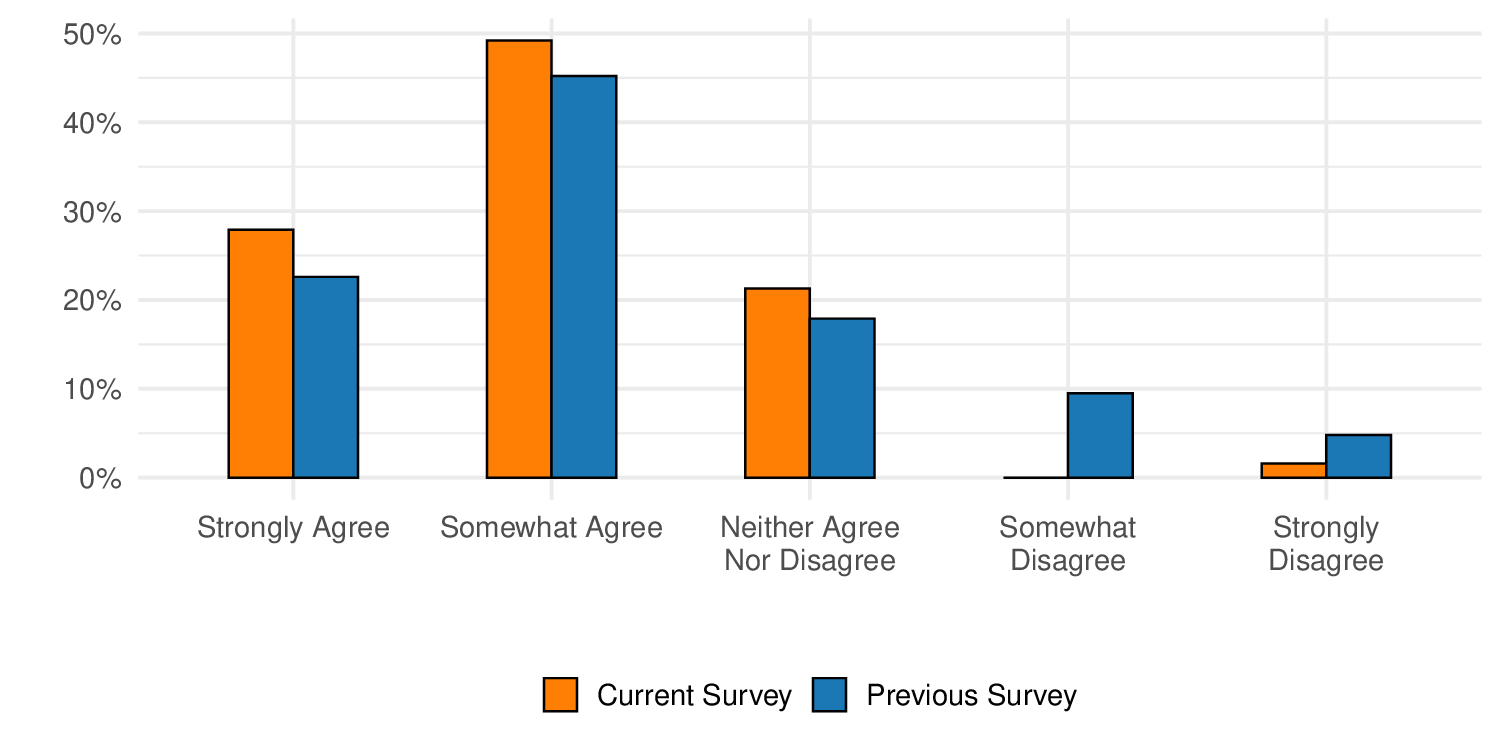}
	\caption{RSEs vs. developers' views on whether peer code review helps decrease code complexity [Q18]}
	\label{fig:Q18}
\end{figure*}

\begin{figure*}[!htb]
	\includegraphics[width=1.0\textwidth]{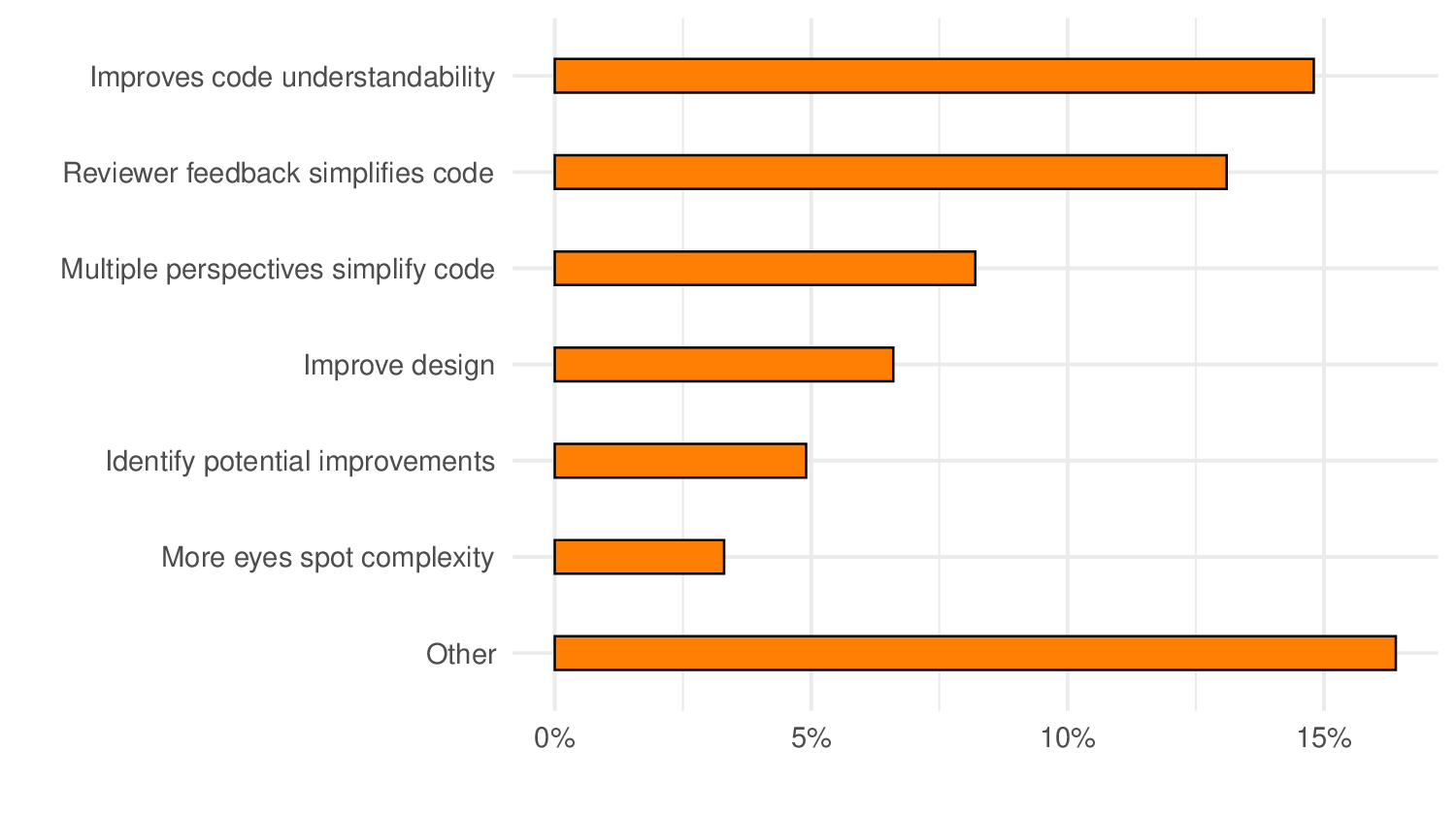}
	\caption{RSEs' positive explanations on peer code review decreasing code complexity [Q19]}
	\label{fig:Q19_pos}
\end{figure*}

The majority of RSEs agree that code reviews help reduce code complexity (Figure~\ref{fig:Q18}). 
This result is similar to the result from the previous survey.
Figure~\ref{fig:Q19_pos} shows the categorized free-response answers explaining why peer code review decreases code complexity. 

\begin{figure*}[!htb]
	\includegraphics[width=1.0\textwidth]{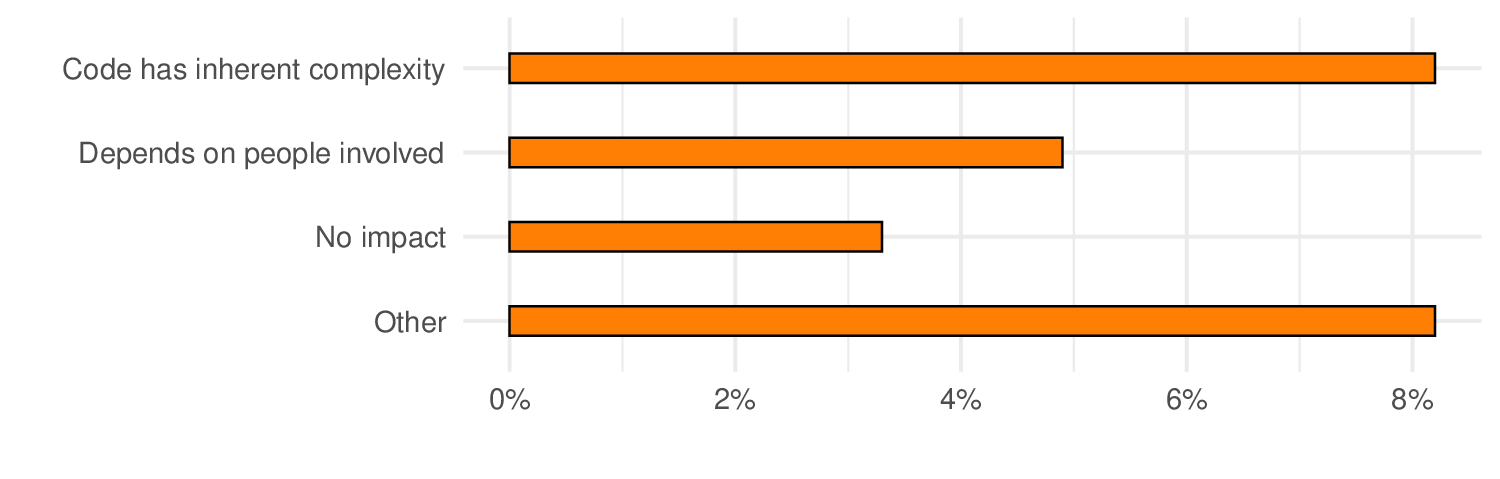}
	\caption{RSEs' negative explanations on peer code review decreasing code complexity [Q19]}
	\label{fig:Q19_neg}
\end{figure*}

Conversely, for those who did not think peer code review reduced code complexity, Figure~\ref{fig:Q19_neg} shows the categorized free-response answers.
One of the primary reasons was that research software has inherent complexity due to the complexity of the problems being solved. 
As one respondent noted, ``Code complexity is not related to peer reviews and rather comes from the nature of the domain.'' 
Another added, ``Software requirements may result in complex code, regardless of review.''

\subsubsection{Theme 4: Peer code review finds performance bottlenecks}
\begin{figure*}[!htb]
	\includegraphics[width=1.0\textwidth]{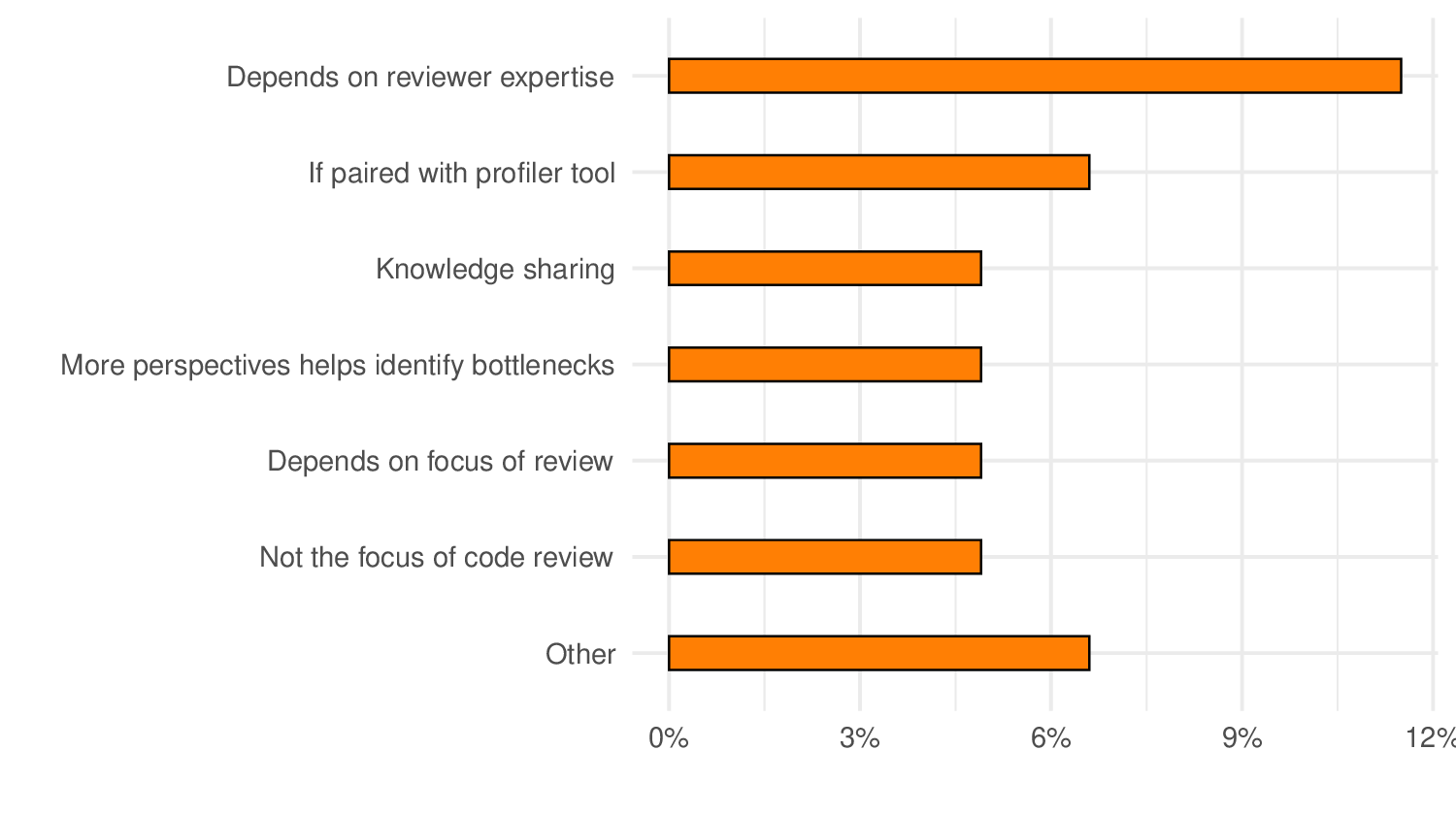}
	\caption{RSEs' positive explanations on how peer code review identifies performance bottlenecks [Q21]}
	\label{fig:Q21_pos}
\end{figure*}

\begin{figure*}[!htb]
	\includegraphics[width=1.0\textwidth]{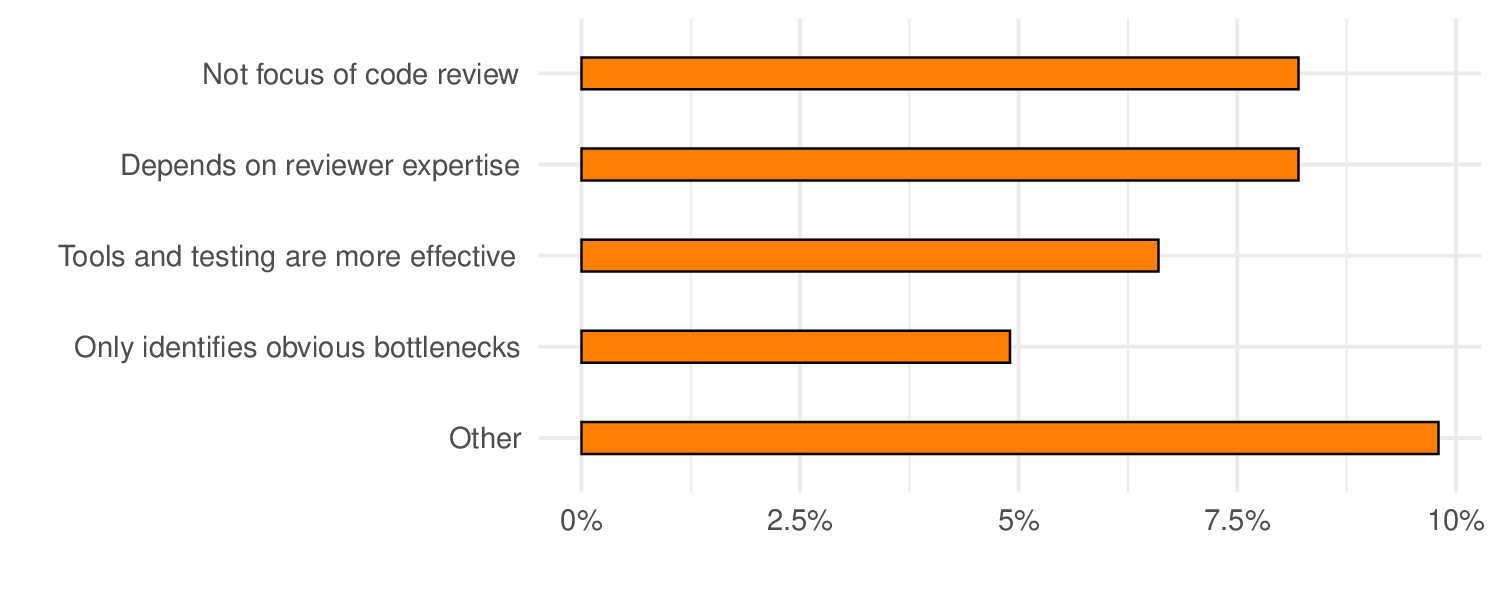}
	\caption{RSEs' negative explanations on peer code review finding performance bottlenecks [Q21]}
	\label{fig:Q21_neg}
\end{figure*}

Just over half of the respondents \textit{strongly agreed} (13\%) or \textit{somewhat agreed} (43\%) that code reviews help identify performance bottlenecks or optimization opportunities. 
Figure~\ref{fig:Q21_pos} and~\ref{fig:Q21_neg} show the results from the free-response question explaining how peer code reviews had either a positive impact on performance bottlenecks or a negative impact on performance bottlenecks.
Across these two figures, some common observations emerge.
First, whether peer code review helps with performance optimization \textit{depends on the reviewer expertise}, as that response appears in both figures. 
Regarding this response, one RSE explained, ``Calling out optimization problems or opportunities occurs regularly, as reviewers are often well-positioned to spot these issues, whereas some developers may not be.''
Second, it is also clear that if reviewers do not think performance is the \textit{focus of the review}, which will limit the types of performance issues identified. 
As one respondent noted, ``It can be the case that a reviewer identifies performance bottlenecks or optimization opportunities. However, in general, I wouldn't consider this to be the primary aim of code review.'' 
Another explained that the performance optimization depends on the focus of code review, ``If performance is a requirement, and review requirements ask for performance reviews or review process includes performance tests, yes. Otherwise, not necessarily.''

\subsubsection{Theme 5: Peer code review reduces bugs in research software}
\begin{figure*}[!htb]
	\includegraphics[width=1.0\textwidth]{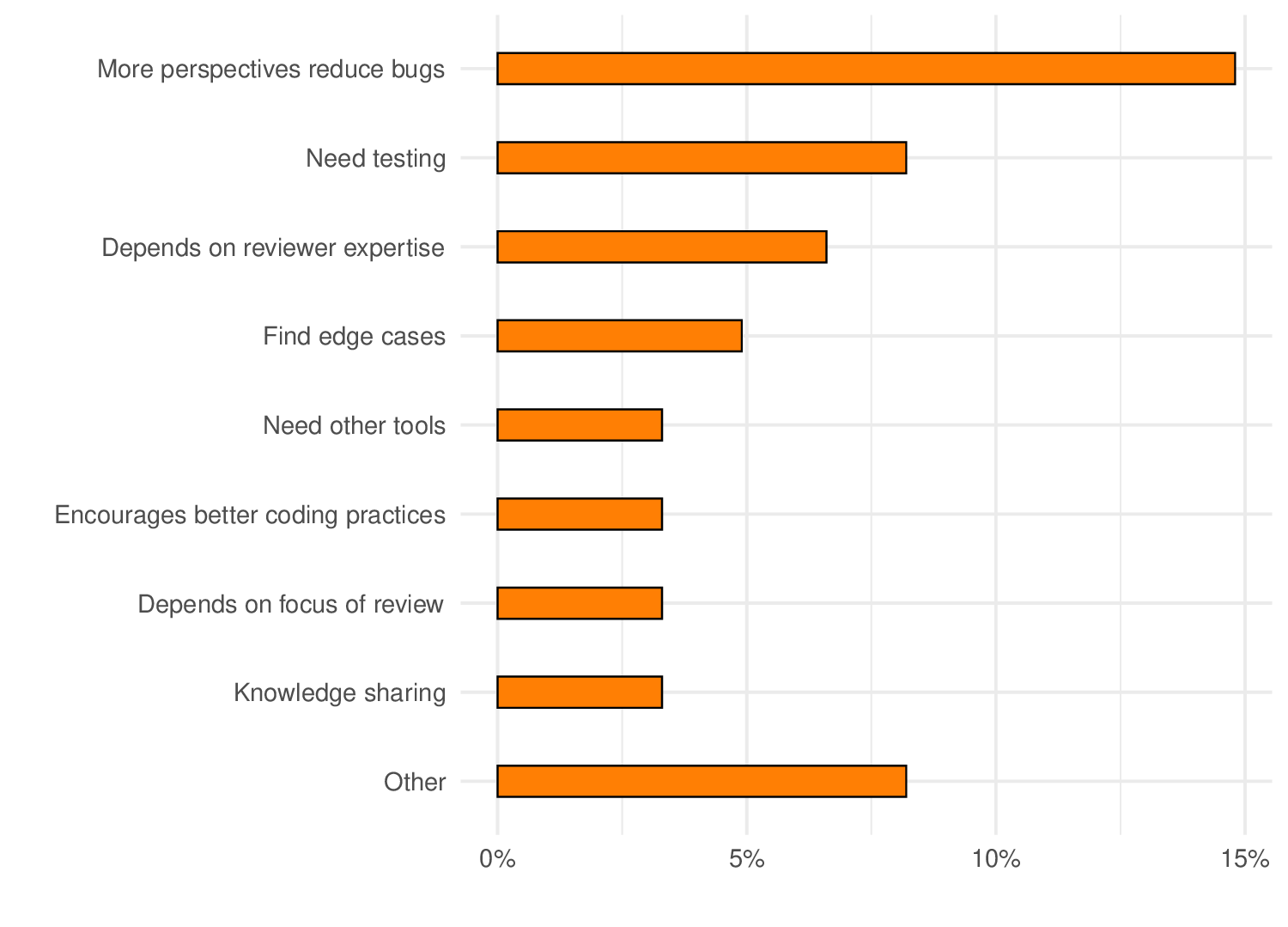}
	\caption{RSEs' positive explanations on peer code review producing more bug-free research software [Q23]}
	\label{fig:Q23_pos}
\end{figure*}

The vast majority of the respondents \textit{strongly agreed} (41\%) or \textit{somewhat agreed} (44\%) that peer code review reduces bugs in research software.
Figure~\ref{fig:Q23_pos} shows the results from the free-response question explaining this answer.
The most common explanations were that \textit{more perspectives reduce bugs}: ``More eyes means more chances for bugs to be caught.'' and that code review is more effective when \textit{paired with testing}: ``Not all bugs will even be found by review. Testing also helps.''

\subsubsection{Theme 6: Peer code review increases maintainability}
\begin{figure*}[!htb]
	\includegraphics[width=1.0\textwidth]{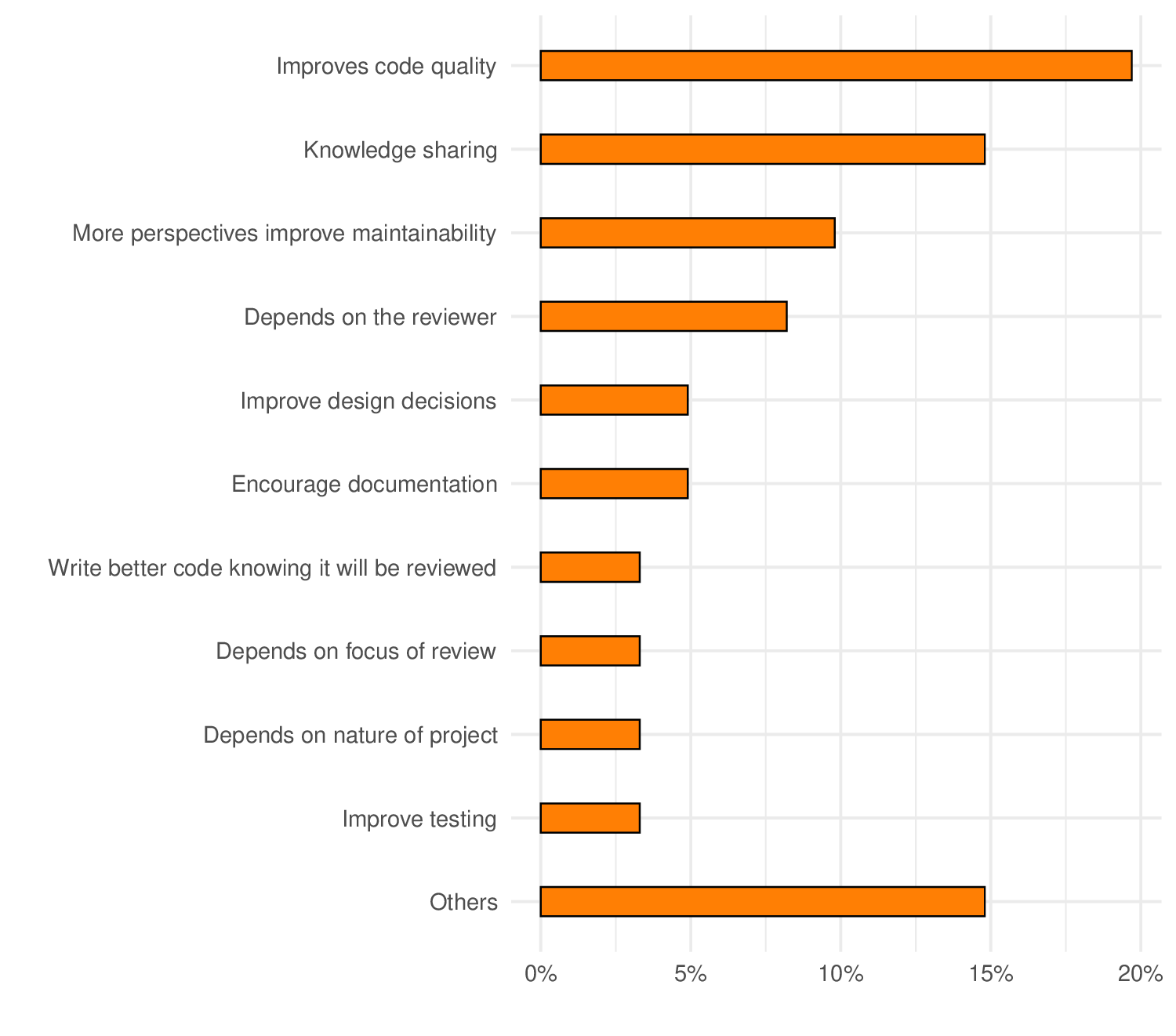}
	\caption{RSEs' explanations on how peer code review increases research software maintainability [Q25]}
	\label{fig:Q25}
\end{figure*}

The vast majority of the respondents \textit{strongly agreed} (62.3\%) or \textit{somewhat agreed} (29.5\%) that peer code review increases maintainability.
Figure~\ref{fig:Q25} shows the results from the free-response question explaining this answer.
The most common answer was that peer code review \textit{improves code quality}. 
Respondents explained this response in several ways, such as making code more readable, clearer, and easier to understand, as well as catching edge cases or encouraging good code. 
One respondent noted, ``In addition to helping ensure the code is readable and understandable by someone other than the author, it exposes at least one other person to the code.'' 
The second most common response focused on \textit{knowledge sharing} as a reason why peer code review helps with maintainability. 
One RSE explained, ``More people know the code, more people understand the code, more people to maintain the code and to streamline.''
However, there are several other responses that have very similar frequencies.
Many of these reasons emphasized that the impact of peer code review on maintainability \textit{depends on the reviewer, the focus of the review, and the nature of the project}. 
As one respondent explained, ``At least it should if at least one reviewer is looking at it from the perspective of long-term sustainability. If that’s not the case, then the maintainability may not be impacted much at all—it all depends on what the reviewers are looking for.'' 
Another echoed this, noting that the outcome often varies ``depending on the experience, expertise, and focus of the reviewers and the nature of the project.''

\subsubsection{Theme 7: Peer code review reduces technical debt}
\begin{figure*}[!htb]
	\includegraphics[width=1.0\textwidth]{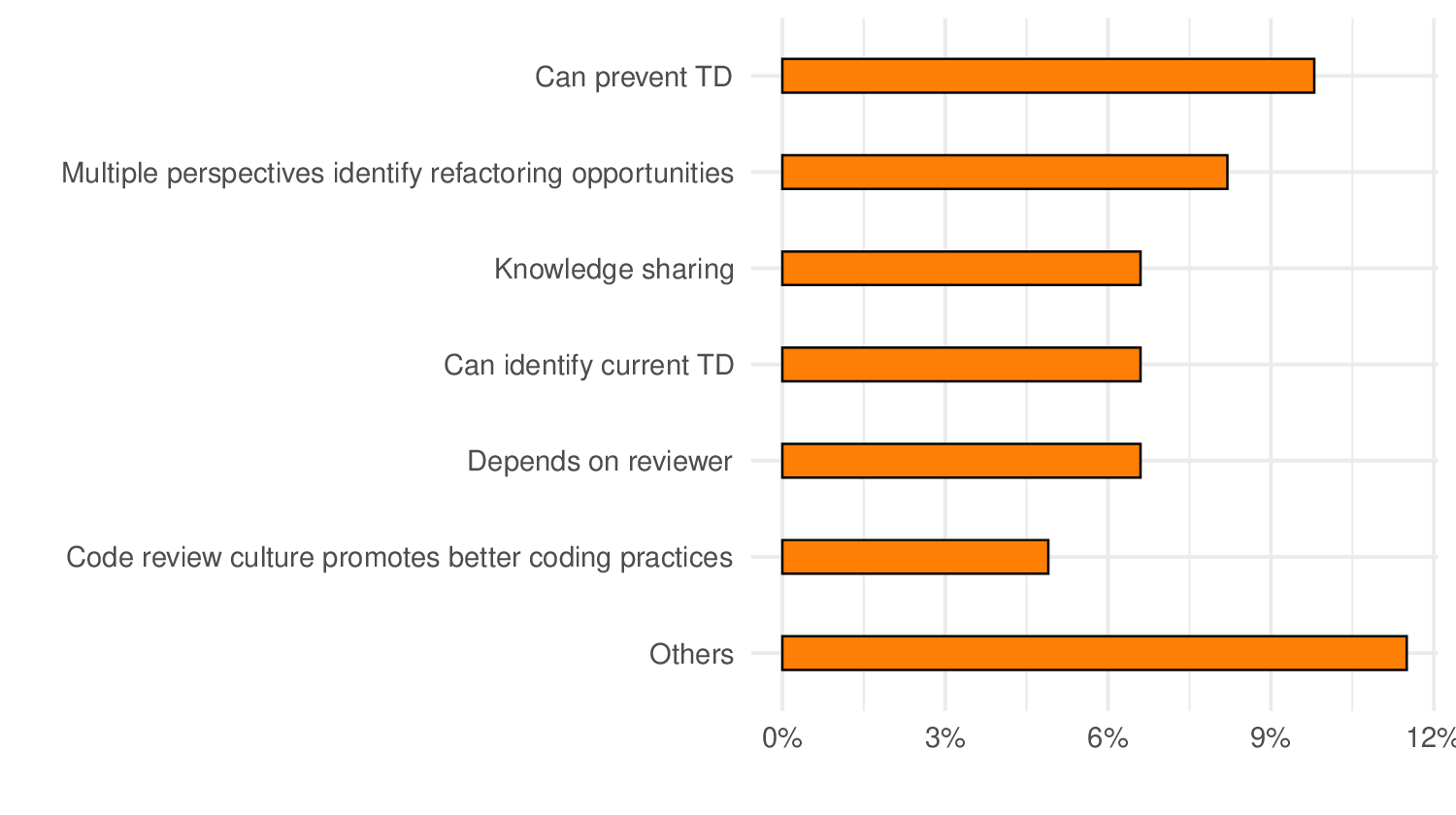}
	\caption{RSEs' positive explanations on how peer code review reduces technical debt [Q27]}
	\label{fig:Q27_pos}
\end{figure*}

The majority of the respondents \textit{strongly agreed} (31\%) or \textit{somewhat agreed} (41\%) that peer code review reduces technical debt.
Figure~\ref{fig:Q27_pos} shows the results from the free-response question explaining this answer.
There was not one dominant answer indicating there are various reasons why respondents believe peer code review reduces technical debt.
One interesting observation is in addition to identifying technical debt, some respondents suggested peer code review actually prevents technical debt. 
As one RSE explained, ``Hopefully it prevents new technical debt from being added to the code base or points out existing technical debt that should be fixed at some point.''

\subsection{RQ3: What difficulties do research software engineers face with peer code review?}

\begin{figure*}[!htb]
	\includegraphics[width=1.0\textwidth]{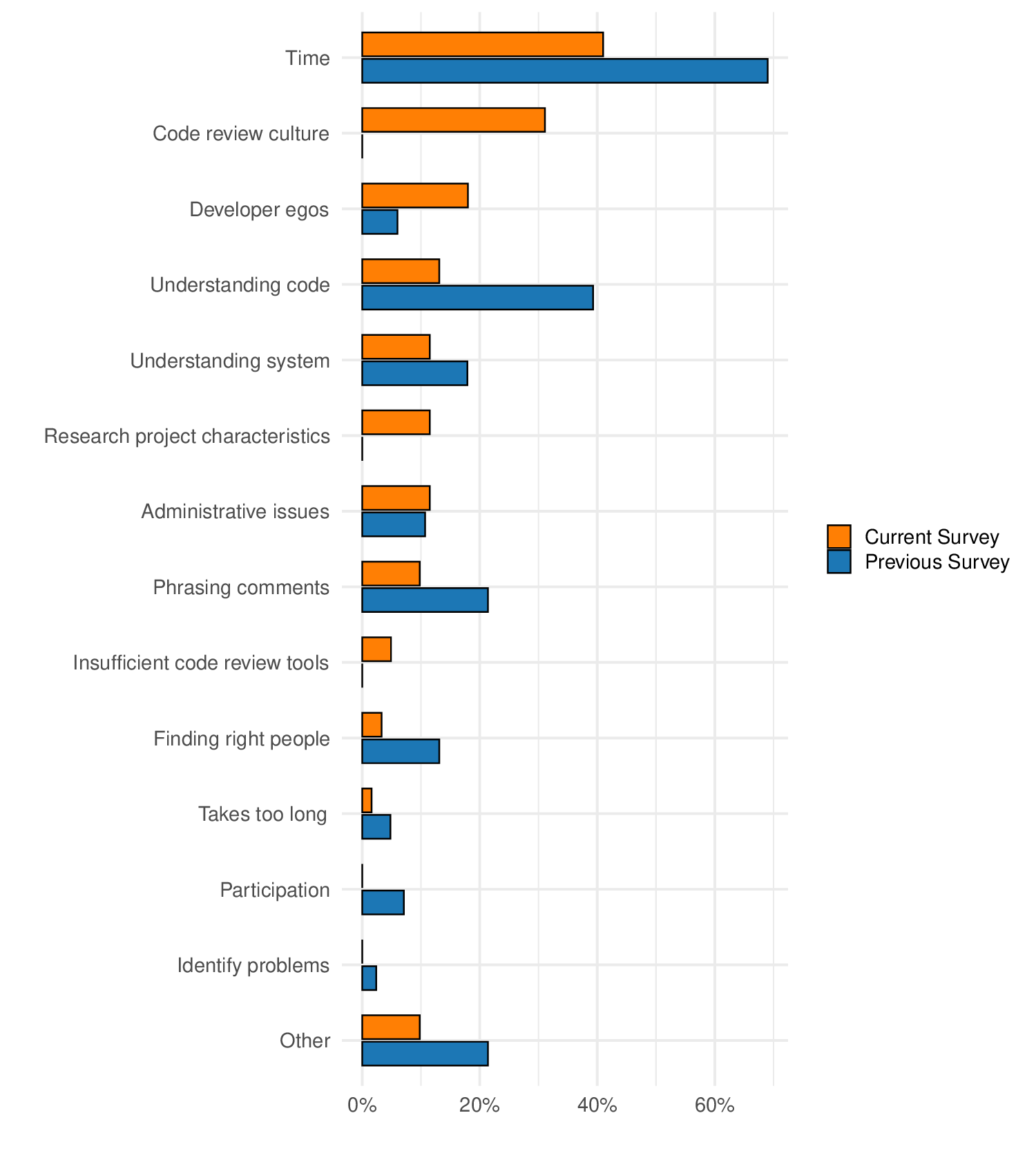}
	\caption{Challenges and barriers in the peer code review process: RSEs vs. developers [Q28]}
	\label{fig:Q28}
\end{figure*}

Figure~\ref{fig:Q28} lists the analyzed results from the free-response question asking about barriers and challenges in the peer code review process.
The most commonly reported challenge RSE's faced was finding \textit{time} to conduct the review.
The second most common challenge was \textit{culture}, which collected ideas about recognition, motivation, and providing a safe space for newcomers.

Comparing these results with those from the previous survey highlights some interesting observations.
First, \textit{time} was the most common barrier for both questions.
Second, the idea of culture was only mentioned by the RSEs and not in the original survey, suggesting that RSEs have a better insight into how the operation of teams may affect the outcomes. 
As one respondent noted, ``Getting people to be comfortable working out in the open where everyone can see the work you do at all times is perhaps the largest cultural barrier.
''
Finally, RSEs mentioned that research projects have some special characteristics, such as a small team, inherent complexity, and lack of funding. 
One RSE mentioned, ``Teams are too small, project scopes are too short term, not enough funding for maintenance of research software.  A lot of wasted effort in isolated research communities''

\subsection{RQ4: What improvements to the peer code review process do research software engineers need?}

\begin{figure*}[!htb]
	\includegraphics[width=1.0\textwidth]{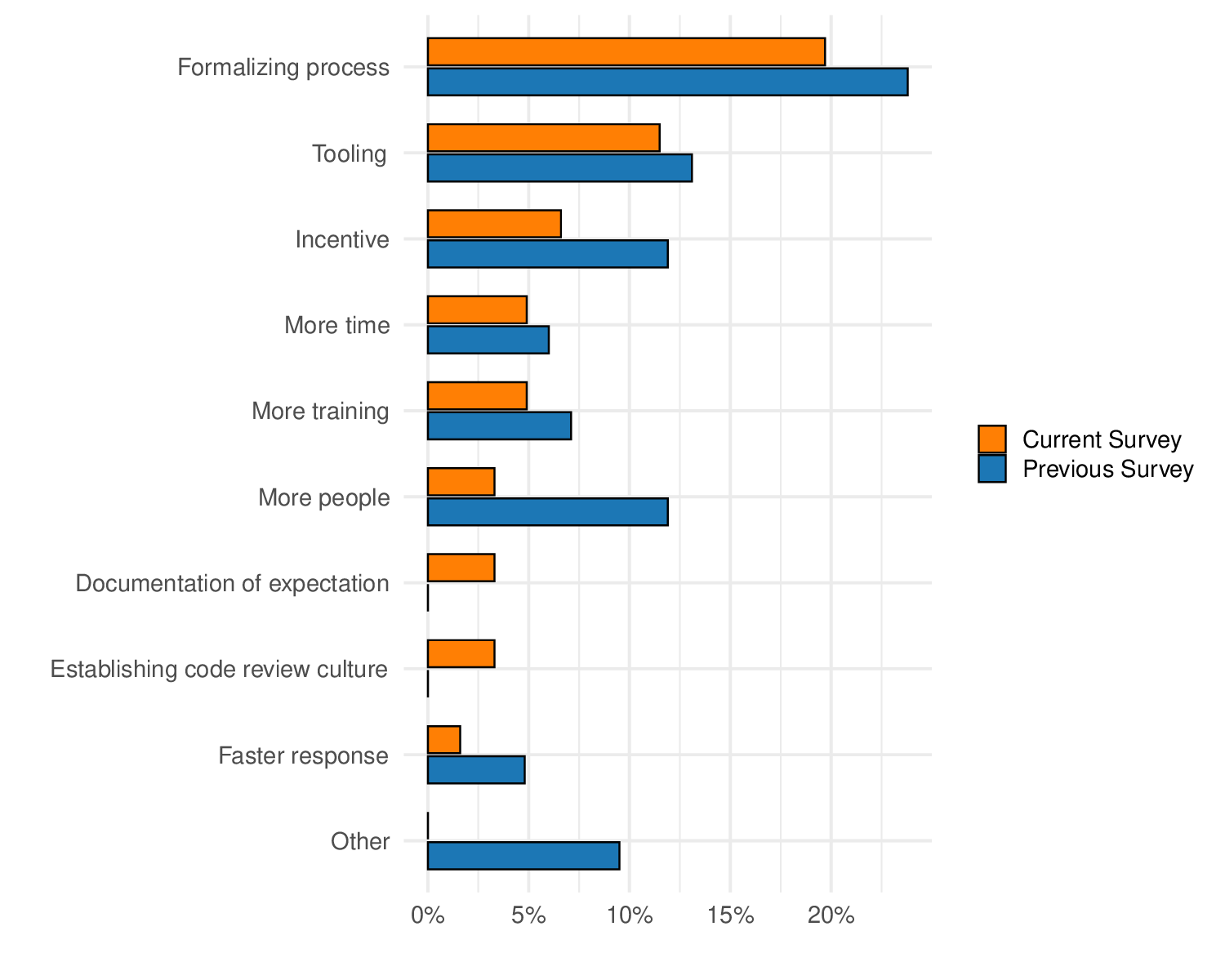}
	\caption{Improvements to the peer code review process: RSEs vs. developers [Q30]}
	\label{fig:Q30}
\end{figure*}

Because Q29 and Q30 produced similar responses, we focus on Q30. 
Figure~\ref{fig:Q30} shows the analyzed results from the free-response question on how to improve the peer code review process.
The most common answer focused on \textit{formalizing the process} with respondents suggesting ``formal checklists of steps a project expects the reviewer to take'' and ``clear guidelines, which we really don’t have, and which take time to develop.''.
The second most common answer, \textit{tooling}, is another way to formalize the process through the addition of tools. 
As one respondent explained, ``Adding as much automation as possible is important—for example, finding and fixing code style inconsistencies automatically allows reviewers to concentrate on more abstract aspects of the code.''

When comparing the results from this survey to those from the previous survey, two items stand out.
The RSEs suggested \textit{documenting expectations} and \textit{culture} as ways to improve the process. 
Respondents highlighted the need for ``documentation of expectations for both code authors and reviewers,'' including clarifying responsibilities such as ``who is responsible for maintaining the code? what happens if a serious bug is found?''. 
Others stressed the importance of culture, noting that ``it should never be used or viewed as a threat.''
These answers are similar in that they focus on ensuring the process is clear and is part of the normal team processes.

\section{Discussion}
\label{sec:Discussion}
In this section, we first summarize the key insights derived from our findings relative to each research question and compare them with those from our previous study on research software developers. Table~\ref{tab:comparison-summary} presents this comparison, highlighting the main similarities and differences across studies.
We then explore the implications of these results. 
\subsection{Answers to Research Questions}

For each research question, we discuss the relevant results.

\subsubsection{RQ1: How do research software engineers perform peer code review?} 

Research software engineers (RSEs) conduct peer code reviews primarily through pull requests on platforms like GitHub or GitLab. One to three individuals typically participate, dedicating about five hours per week. Responses to review comments are generally received within a week, reflecting an efficient workflow.

Reviews focus on identifying code mistakes and design issues, prioritizing correctness. Interactions among teammates are usually positive, with mentoring considered an important aspect of the process.

Positive experiences include improved code quality, a structured process, and knowledge sharing. 
Negative experiences include unhelpful reviews with harsh comments, disagreements, poor processes, and taking too long.

Compared to the previous study, we observe that the dominant use of pull requests remains consistent. 
However, the RSEs typically involve fewer reviewers in code review. 
We also identified a new factor that influences the acceptance of review requests: the impact of the change.
While the time spent on reviews is similar, the initial responses take slightly longer on average, but the final decisions are reached more quickly. 
Overall, the RSEs reported a broader range of both positive and negative experiences with code review compared to the previous survey.

\vspace{8pt}
\noindent
\subsubsection{RQ2: What effect does peer code review have on research software?}

Peer code review significantly enhances research software by improving quality, maintainability, and effectiveness. 
The most frequently reported benefits are better code quality and increased knowledge sharing among team members.

A primary advantage of peer code review is its ability to improve correctness. 
While many agree that it decreases code complexity, some note that the inherent complexity of research software, stemming from the problems it addresses, can limit this benefit.

Peer code review also aids in identifying performance bottlenecks and is credited with increasing maintainability by fostering a shared understanding of the codebase, which helps reduce technical debt over time.

Overall, RSEs regard peer code review as a vital practice for improving software quality, supporting collaboration, and promoting knowledge exchange within research teams.

The result of this survey found similar benefits for peer code review as we found in the first survey.
A similar proportion of respondents strongly agreed on its importance, with improving code quality and knowledge sharing remaining the top reasons. 
Most respondents from both surveys agreed that peer code review improved correctness and decreased code complexity. 
In this study, we also identified additional ways in which peer code review can help such as, reducing bugs, increasing maintainability, detecting performance bottlenecks depending on the reviewer and the review’s focus, and identifying and preventing technical debt.

\vspace{8pt}
\noindent
\subsubsection{RQ3: What difficulties do research software engineers face with peer code review?}

RSEs face distinct challenges in peer code review. 
Time constraints are a major issue, as effective reviews require significant effort.

Cultural barriers also affect the process, as reviews are sometimes undervalued and lack sufficient team support. 
Specific project characteristics, such as small team sizes and the complexity of research tasks, further complicate the review process.

RSEs also reported gaps in peer review tools, noting that existing tools often lack advanced features like static code analysis.
Addressing these challenges requires improvements in tool support and team practices to better align with the needs of research software development.

We found that the main challenge was consistent between the two studies. 
Both groups of respondents cite a lack of time as the primary barrier to peer code review. 
In this study, we saw the absence of code review culture was a major obstacle. 
Conversely, in the previous study, code review participation was an obstacle; however, it was not mentioned in this study.

\vspace{8pt}
\noindent
\subsubsection{RQ4: What improvements to the peer code review process do research software engineers need?}

RSEs suggest several improvements to the peer code review process, primarily the formalization of the review process, better tooling, more time for reviews, and additional training. 
Formalization includes clear guidelines and quality checks, while improved tooling would automate repetitive tasks. More time for thorough reviews and training for new team members are also emphasized as key improvements.

Similar to the previous study, formalizing the process remained a key improvement. 
In this study, respondents identified new concepts that positively impact peer code review, including documenting expectations and establishing a code review culture.

\begin{table}[htb]
\centering 
\caption{Comparison of key findings between the prior study and this study.}
\label{tab:comparison-summary}
\begin{tabular}{|p{.05\textwidth}|p{.2\textwidth}|p{.13\textwidth}|p{0.45\textwidth}|}
\hline
\textbf{RQs} & \textbf{Topic} & \textbf{Similarity} & \textbf{Comment}
\\ \hline

RQ1 & Use of pull requests & Similarity & Both studies report pull requests as the main mechanism for peer code review. \\ \hline

RQ1 & Number of reviewers involved & Dissimilarity & RSEs involve fewer reviewers per review compared to the previous study. \\ \hline

RQ1 & Factors influencing review acceptance & Dissimilarity & RSEs identified a new factor influencing review acceptance: the impact of the change. \\ \hline

RQ1 & Response and closure time & Dissimilarity & RSEs report slightly slower initial responses but faster final decisions. \\ \hline

RQ1 & Positive review experiences & Similarity & Both studies highlight improved code quality and knowledge sharing. \\ \hline

RQ2 & Additional review benefits & Dissimilarity & RSEs add benefits like identifying performance bottlenecks, reducing bugs, increasing maintainability, and preventing technical debt. \\ \hline

RQ3 & Top barrier: lack of time & Similarity & Lack of time is consistently cited as the biggest challenge across both studies. \\ \hline

RQ3 & Cultural challenges & Dissimilarity & RSEs mention lack of code review culture; developers previously noted lack of participation. \\ \hline

RQ4 & Need to formalize review process & Similarity & Formalization remains a top recommendation in both studies. \\ \hline

RQ4 & Emphasis on documenting expectations & Dissimilarity & Only RSEs emphasize documenting expectations and establishing review culture. \\ \hline

\end{tabular}
\end{table}

\subsection{Practical Implications}

\noindent
The findings from this study on how RSEs engage in peer code reviews, the effects of such reviews, the challenges faced, and potential improvements have several practical implications for the field of research software.

\begin{itemize}
    \item \textit{Adoption of Structured Processes.} Formalizing the peer code review process can improve consistency, efficiency, and overall quality. Research teams can benefit from developing clear guidelines, using retrospectives for feedback, and establishing quality benchmarks tailored to the unique challenges of research software development. Such structures would also address administrative gaps, ensuring reviews are conducted thoroughly and systematically.

    \item \textit{Integration of Effective Tools.}
    Tool-based reviews are often underused, presenting an opportunity to automate repetitive tasks and improve bug detection. Automation should address tasks like fixing code style inconsistencies, enabling reviewers to focus on higher-level aspects. Additionally, using CI and merging tools for asynchronous reviews, along with AI tools for support, can further streamline the process. Tools should also notify reviewers and reviewees if a pull request remains open for too long, helping to prevent delays in busy projects.

    \item \textit{Enhancing Reviewer Expertise and Collaboration.} Teams should prioritize training initiatives to build both domain-specific and coding expertise among reviewers. Mentorship, already valued by RSEs, can be leveraged further to bridge knowledge gaps and foster collaboration, particularly in small teams where finding knowledgeable reviewers is challenging. Encouraging a culture of psychological safety and mutual respect can enhance communication, improve feedback clarity, and reduce interpersonal challenges during reviews.

    \item \textit{Emphasizing Documentation and Coding Standards.} Clear documentation of review expectations and coding standards can mitigate ambiguity and support reviewers in delivering constructive feedback. This approach aligns with the need for consistent practices and fosters a culture where review outcomes align with broader project goals, including maintainability and knowledge sharing.

    \item \textit{Ensuring Time and Recognition for Effective Reviews.} Given the time-intensive nature of effective reviews, organizations should allocate dedicated time for this process within project schedules. Reviewers should be recognized (e.g., cited as authors), and ownership should be ensured by core developers.

    \item \textit{Focus on Code Maintainability and Collaboration.} Peer reviews improve maintainability and reduce technical debt, but their full potential is realized when paired with other practices like testing and optimization tools. Research organizations should consider peer reviews as part of a broader strategy for ensuring long-term software sustainability and team knowledge-sharing, particularly for research software projects that may transition between teams or evolve over time.
\end{itemize}
 
\section{Threats to Validity}
\label{sec:Threats}

This section outlines potential threats to the validity of our study.

\subsection{Internal Threats}
One potential threat arises from the use of self-reported data, which relies on the perceptions and recall of participants. 
Respondents may not accurately remember their practices or may provide responses they believe to be socially desirable. 
To mitigate this threat, we reused survey questions from the previous survey. 

Additionally, our approach to coding qualitative responses involved multiple authors, reducing the risk of individual bias in interpretation. Any disagreements in coding were resolved through discussion or adjudicated by a third author.

\subsection{External Threats}
Our previous study did not capture information about whether participants identify as RSEs. 
Therefore, we are not able to determine whether any of the respondents in our current survey also participated in the prior survey. 
There may be some overlap, meaning the two datasets may not represent entirely separate groups. 
As a result, some of the results could be influenced by people who participated in both surveys.
However, given the differences in the results, we do not think this threat is significant.

Next, our survey recruitment strategy relied on newsletters, Slack communities, and social media, which may have skewed participation toward individuals interested in code review. 
This recruitment bias could mean our respondents are not fully representative of the broader target population.

Moreover, because we cannot estimate the size of the population that received the survey invitation, we cannot assess response rates or compare the demographics of respondents to the overall population of RSEs and research software developers. 
This threat limits our ability to generalize the findings beyond our sample.

Additionally, our use of social media (e.g., Twitter/X) and community channels to disseminate the survey may have introduced further sampling bias. 
To ensure our respondents were RSEs and represented a different population than the original study, we mitigated this issue by applying a condition: participants must currently work on a research software project to be eligible to participate in the survey. 
Additionally, in our solicitation, we explicitly indicated the population (RSEs) who should participate.

\subsection{Construct Threats}
One potential threat is that participants may misinterpret survey questions, especially given the technical nature of software engineering terms. 
To reduce this threat, we reused the survey questions from the previous survey.

Additionally, while most survey questions remained consistent with our previous study to allow for comparisons, we added new questions to gather targeted insights from RSEs. 
This change could introduce differences in interpretation between the two populations, potentially affecting the comparability of results. 
\section{Conclusion}
\label{sec:Conclusion}
In this paper, we present insights into peer code review practices in research software based on responses from 61 survey participants. 

Building on our prior study of research software developers, this study focuses specifically on Research Software Engineers (RSEs)—professionals who combine domain knowledge with formal software engineering expertise. 
This replication was necessary to determine whether the findings from the earlier study also apply to this more experienced group or whether new challenges emerge when engineering practices are more established. 
Understanding these differences is critical because RSEs often lead or influence code review practices within research teams.

The results of this study extend the earlier findings in meaningful ways. 
While the previous study showed that research software developers value peer code review but use it inconsistently, this replication reveals that many of the same barriers—such as time pressure and unclear expectations—persist even among trained professionals. 
At the same time, RSEs identified additional organizational and cultural barriers, including the absence of a defined review culture and limited institutional support. 
These insights clarify that the main obstacle to broader adoption is no longer a lack of awareness or training, but rather the need for stronger structural and cultural support within research environments.

To provide an overview of the findings from this study, here we briefly summarize how our findings answer each research question:

\vspace{8pt}
\noindent
\textbf{\RQa} \newline RSEs primarily use pull requests and informal communication for code review, often with more structure than general research software developers.  

\vspace{8pt}
\noindent
\textbf{\RQb} \newline RSEs view peer code review as valuable for maintainability, correctness, and knowledge sharing, but cite time pressure, misaligned expectations, and project-specific constraints as key barriers.  

\vspace{8pt}
\noindent
\textbf{\RQc} \newline RSEs recommend process formalization and better documentation but highlight a need for stronger team buy-in and clearer review expectations. 

\vspace{8pt}
\noindent
\textbf{\RQd} \newline While many RSEs are aware of best practices, inconsistent implementation suggests that training alone is insufficient without supportive review culture and infrastructure.

\vspace{8pt}
These findings allow us to draw conclusions that were not possible from the first study.
They show that improving peer code review in research software projects requires not only educating developers but also embedding review into the organizational culture, supported by policies, expectations, and management buy-in. 
By focusing on RSEs, this study provides a clearer picture of how technical expertise interacts with social and structural factors to shape the success of peer code review in research software. 
Together, these results point toward a path for building more sustainable, high-quality research software through a balance of training, culture, and institutional support.

\section*{Acknowledgments}
We thank the survey participants for their time.
We acknowledge support from the Alfred P. Sloan Foundation grant 2022-19559.

\section*{Data Availability Statement}
Due to IRB restrictions, we cannot share the raw data.
Summarized data is available upon request.

\section*{Compliance With Ethical Standards}
\textit{Conflict of Interest:} None \\

\noindent
\textit{Funding:} This work was supported by a grant from the Alfred P. Sloan Foundation. \\

\noindent
\textit{Ethical approval:} This study received IRB approval from the University of Alabama (IRB\#23-09-6940). \\

\noindent
\textit{Informed consent:} Participants had to complete an informed consent form before beginning the survey. \\

\noindent
\textit{Author Contributions:}
All authors contributed to the conception and design of the study.
All authors participated in the data analysis process.
All authors contributed to the manuscript. 
All authors read and approved the final manuscript. 
\bibliographystyle{spbasic}       
\bibliography{references_Nasir}   

\begin{thebibliography}{19}
\providecommand{\natexlab}[1]{#1}
\providecommand{\url}[1]{{#1}}
\providecommand{\urlprefix}{URL }
\expandafter\ifx\csname urlstyle\endcsname\relax
  \providecommand{\doi}[1]{DOI~\discretionary{}{}{}#1}\else
  \providecommand{\doi}{DOI~\discretionary{}{}{}\begingroup
  \urlstyle{rm}\Url}\fi
\providecommand{\eprint}[2][]{\url{#2}}

\bibitem[{AlNoamany and Borghi(2018)}]{7}
AlNoamany Y, Borghi J (2018) Towards computational reproducibility: researcher
  perspectives on the use and sharing of software. PeerJ Computer Science 4,
  \doi{10.7717/peerj-cs.163}

\bibitem[{Anzt et~al.(2020)Anzt, Bach, Druskat, Loffler, Loewe, Renard,
  Seemann, Struck, Achhammer, Aggarwal, Appel, Bader, Brusch, Busse,
  Chourdakis, Dabrowski, Ebert, Flemisch, Friedl, Fritzsch, Funk, Gast, Goth,
  Grad, Hermann, Hohmann, Janosch, Kutra, Linxweiler, Muth, Peters-Kottig,
  Rack, Raters, Rave, Reina, Reißig, Ropinski, Schaarschmidt, Seibold, Thiele,
  Uekerman, Unger, and Weeber}]{5}
Anzt H, Bach F, Druskat S, Loffler F, Loewe A, Renard B, Seemann G, Struck A,
  Achhammer E, Aggarwal P, Appel F, Bader M, Brusch L, Busse C, Chourdakis G,
  Dabrowski PW, Ebert P, Flemisch B, Friedl S, Fritzsch B, Funk MD, Gast V,
  Goth F, Grad J, Hermann S, Hohmann F, Janosch S, Kutra D, Linxweiler J, Muth
  T, Peters-Kottig W, Rack F, Raters F, Rave S, Reina G, Reißig M, Ropinski T,
  Schaarschmidt J, Seibold H, Thiele J, Uekerman B, Unger S, Weeber R (2020) An
  environment for sustainable research software in germany and beyond: current
  state, open challenges, and call for action. F1000Research 9,
  \doi{10.12688/f1000research.23224.1}

\bibitem[{Barker et~al.(2020)Barker, Katz, and Gonzalez-Beltran}]{Barker2020}
Barker M, Katz DS, Gonzalez-Beltran A (2020) Evidence for the importance of
  research software. Zenodo \doi{10.5281/zenodo.3884311},
  \urlprefix\url{https://doi.org/10.5281/zenodo.3884311}, accessed: 2024-11-18

\bibitem[{Barker et~al.(2022)Barker, Hong, Katz, Lamprecht, Martinez-Ortiz,
  Psomopoulos, Harrow, Castro, Gruenpeter, Martínez, and Honeyman}]{3}
Barker M, Hong NCC, Katz D, Lamprecht AL, Martinez-Ortiz C, Psomopoulos F,
  Harrow J, Castro LJ, Gruenpeter M, Martínez P, Honeyman T (2022) Introducing
  the fair principles for research software. Scientific Data 9,
  \doi{10.1038/s41597-022-01710-x}

\bibitem[{Carver et~al.(2022)Carver, Weber, Ram, Gesing, and Katz}]{4}
Carver JC, Weber N, Ram K, Gesing S, Katz D (2022) A survey of the state of the
  practice for research software in the united states. PeerJ Computer Science
  8, \doi{10.7717/peerj-cs.963}

\bibitem[{Combemale et~al.(2023{\natexlab{a}})Combemale, Gray, and
  Rumpe}]{combemale2023research}
Combemale B, Gray J, Rumpe B (2023{\natexlab{a}}) Research software engineering
  and the importance of scientific models. Software and Systems Modeling
  22(4):1081--1083

\bibitem[{Combemale et~al.(2023{\natexlab{b}})Combemale, Gray, and Rumpe}]{1}
Combemale B, Gray J, Rumpe B (2023{\natexlab{b}}) Research software engineering
  and the importance of scientific models. Software and Systems Modeling
  22:1081--1083, \doi{10.1007/s10270-023-01119-z}

\bibitem[{Eisty and Carver(2022)}]{Eisty-Carver:2022}
Eisty N, Carver J (2022) Developers perception of peer code review in research
  software development. Empirical Software Engineering 27

\bibitem[{Eisty et~al.(2025)Eisty, Kanewala, and Carver}]{9}
Eisty NU, Kanewala U, Carver JC (2025) Testing research software: An in-depth
  survey of practices, methods, and tools. ArXiv abs/2501.17739,
  \doi{10.48550/arXiv.2501.17739}

\bibitem[{Goble(2014)}]{10}
Goble C (2014) Better software, better research. IEEE Internet Comput 18:4--8,
  \doi{10.1109/MIC.2014.88}

\bibitem[{Gomez-Diaz and Recio(2022)}]{2}
Gomez-Diaz T, Recio T (2022) Research software vs. research data i: Towards a
  research data definition in the open science context. F1000Research 11,
  \doi{10.12688/f1000research.78195.2}

\bibitem[{Goth et~al.(2023)Goth, Alves, Braun, Castro, Chourdakis, Christ,
  Cohen, Erxleben, Grad, Hagdorn, Hodges, Juckeland, Kempf, Lamprecht,
  Linxweiler, Schwarzmeier, Seibold, Thiele, Waldow, and Wittke}]{11}
Goth F, Alves R, Braun M, Castro L, Chourdakis G, Christ S, Cohen J, Erxleben
  F, Grad J, Hagdorn M, Hodges T, Juckeland G, Kempf D, Lamprecht A, Linxweiler
  J, Schwarzmeier M, Seibold H, Thiele J, Waldow H, Wittke S (2023)
  Foundational competencies and responsibilities of a research software
  engineer. arXiv abs/2311.11457, \doi{10.48550/arXiv.2311.11457},
  \urlprefix\url{https://doi.org/10.48550/arXiv.2311.11457}

\bibitem[{Gruenpeter et~al.(2021)Gruenpeter, Katz, Lamprecht, Honeyman, Garijo,
  Struck, Niehues, Martinez, Castro, Rabemanantsoa, Chue~Hong, Martinez-Ortiz,
  Sesink, Liffers, Fouilloux, Erdmann, Peroni, Martinez~Lavanchy, Todorov, and
  Sinha}]{gruenpeter-etal}
Gruenpeter M, Katz DS, Lamprecht AL, Honeyman T, Garijo D, Struck A, Niehues A,
  Martinez PA, Castro LJ, Rabemanantsoa T, Chue~Hong NP, Martinez-Ortiz C,
  Sesink L, Liffers M, Fouilloux AC, Erdmann C, Peroni S, Martinez~Lavanchy P,
  Todorov I, Sinha M (2021) Defining research software: a controversial
  discussion. \doi{10.5281/zenodo.5504016},
  \urlprefix\url{https://doi.org/10.5281/zenodo.5504016}

\bibitem[{Jacob(2016)}]{8}
Jacob CR (2016) How open is commercial scientific software? The journal of
  physical chemistry letters 7 2:351--3, \doi{10.1021/acs.jpclett.5b02609}

\bibitem[{Leng(2024)}]{34}
Leng J (2024) Do research software engineers have research methods? Open Access
  Government \doi{10.56367/oag-042-10687},
  \urlprefix\url{https://doi.org/10.56367/oag-042-10687}

\bibitem[{Petre and Wilson(2014)}]{article159}
Petre M, Wilson G (2014) Code review for and by scientists. In: Proc. 2nd
  Workshop on Sustainable Software for Science: Practice and Experience

\bibitem[{{Rinku Gupta}(2022)}]{ANL_RSE_Career}
{Rinku Gupta} (2022) Choosing research software engineering as a career path.
  \urlprefix\url{https://www.anl.gov/mcs/article/choosing-research-software-engineering-as-a-career-path},
  accessed: 2024-11-18

\bibitem[{Sochat et~al.(2022)Sochat, May, Cosden, Martinez-Ortiz, and
  Bartholomew}]{sochat2022research}
Sochat V, May N, Cosden I, Martinez-Ortiz C, Bartholomew S (2022) The research
  software encyclopedia: a community framework to define research software.
  Journal of Open Research Software

\bibitem[{Turzo and Bosu(2024)}]{turzo2024makes}
Turzo AK, Bosu A (2024) What makes a code review useful to opendev developers?
  an empirical investigation. Empirical Software Engineering 29(1):6

\end{thebibliography}

\end{document}